\documentclass[iop]{emulateapj}
\usepackage{graphicx,epsfig,epsf}

\usepackage{graphicx}
\usepackage{times}
\usepackage{url}

\makeatletter
\def\url@leostyle{%
\@ifundefined{selectfont}{\def\UrlFont{\sf}}{\def\UrlFont{\small\ttfamily}}}
\makeatother
\bibpunct[;]{(}{)}{;}{a}{}{,}

\def\Vlsr {\ifmmode {V_{\rm LSR}} \else {$V_{\rm LSR}$} \fi}
\def\Ro   {\ifmmode {R_0} \else {$R_0$} \fi}
\def\To   {\ifmmode {\Theta_0} \else {$\Theta_0$} \fi}

\def\simless{\lower2pt\hbox{$\buildrel {\scriptstyle <}
   \over {\scriptstyle\sim}$}}
\def\pd  {\lower0pt\hbox{$\buildrel {^\circ} \over {.}$}}
\def\msun{$M_{\odot}$}

\def\aap{A\&A}

\def\aj{AJ}
\def\apj{ApJ}
\def\apjl{ApJ}
\def\apjs{ApJS}

\def\araa{ARA\&A}

\def\mnras{MNRAS}
  
\def\nat{{Nature}}

\def\pasj{PASJ}
\def\pasp{PASP}
\def\actaa{Acta Astronomica}

\newcounter{lastnote}

\begin{document} 
\title{\sc The Extreme Spin of the Black Hole in Cygnus X-1}

\shorttitle{The Extreme Spin of Cygnus X-1}
\shortauthors{Gou et al.}

\author{Lijun~Gou,$^{1}$  Jeffrey E. McClintock,$^{1}$ Mark J. Reid,$^{1}$
  Jerome A. Orosz,$^{2}$ James F. Steiner,$^{1}$ Ramesh Narayan,$^{1}$
 Jingen Xiang,$^{1}$ Ronald A. Remillard,$^{3}$ 
 Keith A. Arnaud,$^{4,5}$ Shane W. Davis$^{6}$}
\altaffiltext{1}{Harvard-Smithsonian Center for Astrophysics,
  Cambridge, MA, 02138, USA}
\altaffiltext{2}{Department of Astronomy, San Diego State University,
  5500 Campanile Drive, San Diego, CA 92182, USA}
\altaffiltext{3}{Kavli Institute for Astrophysics and Space Research, MIT, 70 Vassar Street, Cambridge, MA 02139, USA}
\altaffiltext{4}{CRESST, NASA Goddard Space Flight Center, 8800 Greenbelt Road, Greenbelt, MD 20771, USA} 
\altaffiltext{5}{Astronomy Department, University of Maryland, College Park, MD 20742, USA}
\altaffiltext{6}{Canadian Institute for Theoretical Astrophysics, University of Toronto, Toronto, ON M5S 3H8, Canada}

\slugcomment{ApJ}
\begin{abstract}

The compact primary in the X-ray binary Cygnus X-1 was the first black
hole to be established via dynamical observations.  We have recently
determined accurate values for its mass and distance, and for the
orbital inclination angle of the binary.  Building on these results,
which are based on our favored (asynchronous) dynamical model, we have
measured the radius of the inner edge of the black hole's accretion
disk by fitting its thermal continuum spectrum to a fully relativistic
model of a thin accretion disk. Assuming that the spin axis of the
black hole is aligned with the orbital angular momentum vector, we
have determined that Cygnus X-1 contains a near-extreme Kerr black
hole with a spin parameter $a_*>0.95$ (3$\sigma$).  For a less
probable (synchronous) dynamical model, we find $a_*>0.92$
(3$\sigma$).  In our analysis, we include the uncertainties in black
hole mass, orbital inclination angle and distance, and we also include
the uncertainty in the calibration of the absolute flux via the Crab.
These four sources of uncertainty totally dominate the error budget.
The uncertainties introduced by the thin-disk model we employ are
particularly small in this case given the extreme spin of the black
hole and the disk's low luminosity.

\end{abstract}

\keywords{accretion, accretion disks -- black hole physics -- stars:
individual (Cygnus X-1) -- X-rays: binaries}

\section{\sc Introduction}

In our two companion papers \citep{rei+2010,oro+2010}, we report
accurate measurements of the distance, black hole mass and orbital
inclination angle for the black-hole binary system Cygnus X-1.
Herein, we use these results to determine the spin of the black hole
primary by fitting the thermal component of its X-ray spectrum to our
implementation of the Novikov-Thorne model\footnote{Note that we have
corrected the original Novikov-Thorne equations \citep{nov+1973} for
the problem identified by \citet{rif+1995}.  The term
``Novikov-Thorne'' here refers to a relativistic and
geometrically-thin accretion disk in Kerr geometry with a no-torque
boundary condition at the disk's inner edge.} of a thin accretion
disk~\citep{lil+2005}.

Cygnus X-1 was discovered at the dawn of X-ray astronomy
\citep{bow+1965} and is one of the brightest and most persistent
celestial X-ray sources. Its compact primary was the first object to
be established as a black hole via dynamical observations
\citep{web+1972,bol+1972}.  For several decades, the source has been
extensively observed at radio, optical, ultraviolet and X-ray wavelengths.

Cygnus X-1 is typically in a hard spectral state, but occasionally it
switches to a soft state, which may persist for up to a year (see
Figure 1).  It is only in this soft state, when the disk spectrum is
prominent, that one can measure the spin using the continuum-fitting
method we employ. In this soft state, which corresponds to the steep
power-law (SPL) state\footnote{Throughout, we use the X-ray states
defined by \citet{rem+2006}: hard; thermal dominant (TD); steep
power-law (SPL); and intermediate states.  In the alternative state
classification scheme of \citet{hom+2005}, the only states reached by
Cygnus X-1 are the Low/Hard, Hard-Intermediate and Soft-Intermediate,
which correspond respectively to our hard, intermediate (i.e.,
hard:SPL) and SPL states.}, a strong Compton component is always
present.  Although Cygnus X-1 has been observed on thousands of
occasions, it has never been observed to reach the thermal dominant
(TD) state, the state that is most favorable for the measurement of
spin via the continuum-fitting method \citep{ste+2009a}.

Following the pioneering work of \citet{zha+1997}, we measure black
hole spin by estimating the inner radius of the accretion disk $R_{\rm
in}$.  In this approach to measuring spin, one identifies $R_{\rm in}$
with the radius of the innermost stable circular orbit $R_{\rm ISCO}$,
which is predicted by general relativity.  $R_{\rm ISCO}$ is a
monotonic function of the dimensionless spin parameter $a_{\rm *}$,
decreasing from 6~GM/c$^2$ to 1~GM/c$^2$ as spin increases from
$a_*=0$ to $a_*=1$ \citep{bar+1972}\footnote {$a_* \equiv cJ/GM^2$
with $|a_*| \le 1$, where $M$ and $J$ are respectively the black hole
mass and angular momentum.}. This relationship between $a_*$ and
$R_{\rm ISCO}$ is the foundation for measuring spin by either the
continuum-fitting method \citep{zha+1997} or by the Fe~K$\alpha$
method \citep{fab+1989,rey+2003}.

The identification of $R_{\rm in}$ with $R_{\rm ISCO}$ is strongly
supported by the abundant empirical evidence that the inner radius of
the disk in the soft state of a black hole binary does not appear to
change even as the temperature and luminosity change.  LMC~X-3 is a
prime example; its inner-disk radius has been shown to be stable to
several percent over a period of 26 years \citep{don+2007,ste+2010a}.
Strong theoretical support for identifying $R_{\rm in}$ with $R_{\rm
ISCO}$ is provided by magnetohydrodynamic simulations of thin
accretion disks, which show the disk emission falling off rapidly
inside the ISCO \citep{sha+2008,rey+2008,pen+2010,kul+2010,nob+2010}.

In our early work on measuring spin, we relied solely on TD-state data
\citep[e.g.,][]{sha+2006,mcc+2006,liu+2008}. More recently, using our
empirical model of Comptonization {\sc simpl} \citep{ste+2009b}, we
have shown that one can obtain values of the inner disk radius using
SPL data that are consistent (within $\sim$5\%) with those obtained
using TD data if the scattering fraction\footnote{$f_{\rm SC}$ is the
fraction of the thermal seed photons that are scattered into the
power-law tail, and it is the normalization parameter of our XSPEC
model {\sc simpl/simplr} (see Section~3.2).}  $f_{\rm SC} \lesssim
25$\% \citep{ste+2009a}.  Consequently, we are now able to routinely
and reliably obtain spin estimates for sources in the SPL state
\citep{gou+2009,ste+2010b} if the scattering fraction is not too
extreme. This development has paved the way for measuring the spin of
Cygnus X-1, which has never been observed in the TD state.

For the continuum-fitting method to succeed, it is essential to have
an accurate value for the black hole mass $M$ in order to express the
radius appropriately as the dimensionless quantity $R_{\rm ISCO}
c^2/GM$.  For the method to succeed, it is furthermore essential to
know the disk luminosity, and therefore to have accurate values for
the distance $D$ and the disk inclination angle $i$, which we infer by
assuming that the orbital angular momentum vector and black-hole spin
axis are aligned (see Section~7.4).  We measured these three critical
parameters in a two-step process: Firstly, using the Very Long
Baseline Array (VLBA), we determined a model-independent distance via
trigonometric parallax that is accurate to 6\%:
$D=1.86_{-0.11}^{+0.12}$~kpc \citep{rei+2010}.  Secondly, using this
accurate distance estimate to constrain the radius of the companion
star, we then modeled an extensive collection of optical data.  For
our favored asynchronous model, Model $\cal D$, we find:
$M=14.8\pm1.0~M_{\sun}$ and $i=27.1\pm0.8$ deg \citep{oro+2010}.  With
the values of these three input parameters in hand, and using three
soft-state X-ray spectra, we determined the spin of Cygnus X-1; our
final value is presented in Section~6.

Spin measurements have previously been obtained using the
continuum-fitting method for eight black holes \citep{mcc+2010}.
Three have low to moderate spins, $a_*~\simless~0.5$; four have high
spins, $a_*\sim0.7-0.9$; and one of them, the archetypal microquasar
GRS~1915+105 \citep{mir+1994,fen+2004}, is a near-extreme Kerr hole
with $a_*>0.98$ \citep{mcc+2006,blu+2009}.

The paper is organized as follows.  In Section 2, we discuss our
observations and the selection and reduction of the X-ray spectral
data.  The analysis of these data and our results are presented
respectively in Sections~3 and 4 for our adopted model, and the
robustness of these results is discussed in Section~5.  (Additional
analysis work and results for several preliminary models are presented
in the appendices.)  In Section~6, we first determine the error in the
spin parameter due to the combined uncertainties in $D$, $M$, $i$ and
the absolute flux calibration, and we then present our final value of
$a_*$ and confidence limits.  Seven distinct topics are addressed in
Section 7, and in Section 8 we offer our conclusions.

\section{Data Selection, Observations and Data Reduction}

There exist very few spectra of Cygnus X-1 that are suitable for the
measurement of the black hole's spin for reasons we now discuss.  A
typical soft-state (and SPL) spectrum is comprised of three principal
elements: a thermal component, a power-law component, and a reflected
component, which includes the Fe~K$\alpha$ emission line.  Three such
spectra are analyzed in detail and illustrated in Section~3.  It is
apparent from an inspection of these spectra that the spectral
coverage must extend to at least 30~keV in order to constrain the
strong power-law and reflection components.  At the same time, because
the temperature of the thermal component is consistently low
($kT\sim0.5$~keV; Appendix~A), one also requires coverage down to
$\approx1$~keV in order to constrain this crucial thermal component,
which is partially absorbed at low energies by intervening gas.  The
rarity of spectra that meet these requirements became clear to us
following our exhaustive search of the thousands of spectra of Cygnus
X-1 that are contained in the HEASARC data archive.  To our surprise,
we found only a single spectrum, SP1 (Table~1), that is suitable for
the measurement of black hole spin\footnote{We decided not to use the
{\it Swift} XRT/BAT spectra, which provide broadband coverage, because
(1) the XRT timing spectra (e.g., {\it Swift} ID 00031651005) show
strong residual features in the energy range 1-3 keV; (2) the Low-Rate
Photodiode spectrum ({\it Swift} ID 00101469000) is poorly calibrated
\citep{rom+2005}; and (3) {\it Swift} provides no coverage at all from
10--15 keV.}.  It was obtained on 30 May 1996 in an observation made
simultaneously using the {\it Advanced Satellite for Cosmology and
Astrophysics (ASCA)} and the {\it Rossi X-ray Timing Explorer (RXTE)}.

There is a second, important reason for the paucity of suitable
spectra, namely that Cygnus X-1 is seldom in the required
disk-dominated state.  This fact is illustrated in Figure~1, which
summarizes the behavior of Cygnus X-1 since 1996 as observed using the
{\it RXTE} All-Sky Monitor (ASM).  As indicated in the figure, we
select only those data for which the spectral hardness SH $<0.7$,
which occurs $<10$\% of the time.  Fortunately, Cygnus X-1 entered its
soft state in mid-2010, and we obtained two additional broadband
spectra on 22 July and 24 July 2010 by making simultaneous
observations using the {\it Chandra X-ray Observatory} and {\it RXTE}
(Table~1).  The times of these two observations and the {\it
ASCA/RXTE} observation are indicated by arrows in Figure~1, and the
corresponding ASM measures of intensity and spectral hardness are
plotted as red stars.  Detailed information on these three
observations is summarized in Table~1.  Three spectra were derived
from these observations: the 1996 archival {\it ASCA/RXTE} spectrum
SP1, and the two {\it Chandra/RXTE} spectra, SP2 and SP3, which were
obtained in 2010.

We first focus on spectrum SP1, which has been described in detail by
others \citep{dot+1997,cui+1998,gie+1999}.  We include in our analysis
all of the data collected by both {\it ASCA} and {\it RXTE}, even
though the observations were strictly simultaneous for only about
1000~s.  This is a reasonable approach because both missions show that
the source intensity was stable during the entire observing period
\citep[see Figure~4 in][]{gie+1999}.  For {\it ASCA}, we consider only
data collected by the second Gas Imaging Spectrometer (GIS) detector,
GIS2 (the GIS3 data were excluded because of an unexplained residual
feature in the spectrum near 1~keV).  We disregard the Solid-state
Imaging Spectrometer (SIS) data because of data-rate limitations.  For
{\it RXTE}, we use all of the data collected by all five Proportional
Counter Array (PCA) detectors.  For the main analysis, we disregard
the {\it RXTE} High Energy X-ray Timing Experiment (HEXTE) data
because the useful bandwidth of the PCA, which extends to 45 keV,
already provides more than adequate energy coverage. We demonstrate
this fact in Section~5.2 where we show that the inclusion of HEXTE
data has no significant affect on our results.

The GIS spectra were extracted following the standard procedures
described in The {\it ASCA} Data Reduction Guide\footnote{The manual
is available at
\url{http://heasarc.gsfc.nasa.gov/docs/asca/abc/abc.html}}.  The GIS2
spectrum was fitted over the energy range 0.7--8.0 keV.  We did not
correct the effective area of the detector because the instrument team
had already done this calibration to an accuracy of 3\% using the
standard Crab spectrum \citep{too+1974,mak+1996,ste+2010a}.

We reduced the {\it RXTE} PCA data following the same procedures
described in \citet{mcc+2006}. Data reduction was performed with
standard tools from the HEASOFT package provided by NASA.  The
critical steps in determining the PCA background and making the
response files for spectral analysis utilized HEASOFT version 6.10.
X-ray spectra were extracted from the Standard 2 telemetry mode, which
provides coverage of the full PCA bandpass every 16 s.  Data from all
Xe gas layers of PCU-2 were combined to make each spectrum. The
background spectrum was determined with the ``bright source'' model.
Redistribution matrix files and ancillary response files were freshly
generated individually for each PCU layer and then combined into a
single response file using the tool \textit{pcarsp}. We used the PCA
response matrices\footnote{For a description of the latest response
files, see
\url{http://www.universe.nasa.gov/xrays/programs/rxte/pca/doc/rmf/pcarmf-11.7}}
v11.7 released on 2009 August 17, which allowed us to obtain reliable
fits over the energy range 2.55--45.0 keV.

We corrected for the effective area of the PCA using the spectrum of
the Crab Nebula as a standard source and using our recently-adopted
and improved method, which is described in \citet{ste+2010a}.
Specifically, we compared the Crab spectrum of \citet{too+1974}
($\Gamma=2.1$ and $N=9.7~{\rm photons~s^{-1}~keV^{-1}}$) to parameters
obtained by analyzing proximate archival observations of the Crab.  In
this way, we determined a pair of correction factors for spectrum SP1:
a normalization correction factor $C_{TS} = 1.123$ (the ratio of the
observed normalization to that of Toor \& Seward) and a correction to
the slope of the power law, $\Delta \Gamma_{TS} = 0.023$ (the
difference between the observed value of $\Gamma$ and that of
\citeauthor{too+1974}).  These corrections were applied in all of our
analysis work via a custom XSPEC multiplicative model {\sc CRABCOR}.
We also corrected the detector count rates for dead time by the factor
1.048.

We turn now to the recent {\it Chandra/RXTE} observations (Table~1;
spectra SP2 and SP3).  The observations performed by the two
spacecraft were mostly simultaneous, overlapping for about 4 ks, while
the total duration of each {\it Chandra} observation was
$\approx6$~ks.  Because the source intensity for both observations was
constant to within 10\%, we included in our analysis all of the {\it
Chandra} data, plus the strictly simultaneous {\it RXTE} PCA data.

Our {\it Chandra} observations were made using the High-Energy
Transmission Grating (HETG) and the Advanced Camera for Imaging and
Spectroscopy (ACIS) \citep{can+2005,gar+2003}.  The data-rate
limitation of the detectors makes it challenging to observe a bright
and variable source like Cygnus~X-1.  The principal problem is
``pile-up,'' i.e., the registering of two or more photons in the same
or adjacent pixels within a single frame time.  Given the
uncertainties, and as a hedge, we performed the pair of {\it Chandra}
observations using two different instrumental configurations, Timed
Exposure (TE) mode and Continuous Clocking (CC) faint mode, which have
contrasting virtues and limitations. 

In reducing the {\it Chandra} TE-mode data, we followed the method
described by~\citet{smi+2002}. In order to avoid saturating the
telemetry, only the data for the High Energy Grating (HEG; -1 order)
and Medium Energy Grating (MEG; +1 order) components of the HETG were
recorded. The data for the readout streak on the same side of the HEG
and MEG spectra were also recorded. We extracted the ``streak'' and
background spectra following the recommended procedures
\footnote{\url{http://cxc.harvard.edu/ciao/threads/streakextract/}}.
Although the net exposure time for the HETG spectrum is 2.15~ksec, the
effective exposure time for the streak spectrum is only about
8~sec\footnote{The procedure for calculating this exposure time can be
found at
http://cxc.harvard.edu/ciao/threads/streakextract/index.html\#exposure}.
For this spectrum, we estimate that less than 3\% of the events are
affected by pile-up, and we therefore use the full 0.5--10~keV
bandwidth.  For the dispersed grating spectrum, we only included data
for which $<5$\% of the events are piled up; for the HEG and MEG
respectively these data are in the energy ranges 0.7--0.9~keV and
7.0--10.0~keV.

In the CC mode, the frame time is reduced to about 3~ms (compared to
1.3 s for the TE mode) by continuously transferring the data from the
image array to the frame-store array.  While this essentially
eliminates pile-up, the details of the spatial distribution in the
columns are then lost due to the collapse of the 2D image into a 1D
image.  Again, only the orders of the HEG and MEG spectra mentioned
above were recorded.  The spectra were extracted following the
standard
procedures\footnote{\url{http://cxc.harvard.edu/ciao/threads/spectra\_hetgacis/}}. We
fitted these data over the full energy range 0.5--10.0~keV, except for
the 1.3--2.0 keV chip gap in the MEG spectrum.  A downside of the CC
mode is that its calibration is less certain than that of the TE mode.

For {\it RXTE}, lacking data from all five Proportional Counter Units
(PCUs), we elected to use only the data from what historically has
been the best-calibrated detector, PCU2.  (One obtains essentially the
same results for the {\it Chandra/RXTE} spectra using any combination
of the available PCUs; likewise, for the {\it ASCA/RXTE} spectrum, it
is unimportant whether one uses all PCUs, as we did, or PCU2 alone.)
As described above, we corrected the effective area of the detector.
For both observations, the normalization correction is $C_{\rm TS} =
1.073$, and the power-law slope correction is $\Delta \Gamma_{\rm TS}
= 0.029$.  The dead time corrections for SP2 and SP3 are respectively
1.052 and 1.044. 

Finally, we included systematic errors in the count rates in each PHA
channel to account for uncertainties in the instrumental responses:
1\% for GIS2 and 0.5\% for all the PCUs. All the {\it Chandra} data
were binned to achieve a minimum number of counts per channel of 200;
no systematic error was included because the statistical error is
large.


\section{Data Analysis}

A typical spectrum of Cygnus X-1 is comprised of three principal
elements: a thermal component, a power-law component, and a reflected
component that includes the Fe~K$\alpha$ emission line.  The
structures in the X-ray source that give rise to these components,
namely the accretion disk and its corona, are illustrated in Figure~2.
The spectral components themselves, in relation to the total observed
spectrum, are shown plotted in Figure~3, which illustrates the results
of the relativistic analysis described in this Section.

The data analysis and model fitting throughout this paper were
performed using XSPEC\footnote{{\sc XSPEC} is available at
\url{http://heasarc.gsfc.nasa.gov/xanadu/xspec/ }} version 12.6.0
\citep{arn+1996} and, unless otherwise indicated, errors are
everywhere reported at the $1\sigma$ level of confidence.  In this
section (and throughout Appendices A and B), we fix the key input
parameters $D$, $M$ and $i$ at their fiducial values, which are given
in Section 1.

\subsection{Seven Preliminary Models}

Our adopted model that is featured below, and upon which all of our
results are based, was constructed by working in detail through a
progression of seven preliminary models.  We now briefly comment on
these models, which are presented in full in Appendices A and B.

{\it Nonrelativistic models:} The central component of our three
nonrelativistic models, Models NR1--NR3, is the familiar
accretion-disk model component {\sc diskbb}\footnote{For descriptions
of the XSPEC models, see\\
\url{http://heasarc.gsfc.nasa.gov/docs/xanadu/xspec/manual/XspecModels.html
}} \citep{mit+1984,mak+1986}, which does not include any relativistic
effects or the effects of spectral hardening, and which has an
inappropriate boundary condition at the disk's inner edge
\citep{zim+2005}.  We nevertheless employed {\sc diskbb} as an
exploratory tool because it has been widely used for decades, and it
therefore allows us to compare our reduction/analysis results (for a
spectrum of interest) to published results.  Furthermore, this
familiar model returns a direct and useful estimate of the temperature
at the inner edge of the disk, which we use in order to make
comparisons between Cygnus X-1 and other black hole binaries.  The
details of this nonrelativistic analysis and the satisfying results
obtained for Model NR3 -- namely consistent values of the inner-disk
radius and temperature for our three spectra -- are presented in
Appendix A.

{\it Relativistic models:} Similarly, in addition to our adopted
relativistic model (Section 3.2), in Appendix B we present our
analysis and results for four relativistic models, Models R1--R4, that
are built around our fully relativistic accretion-disk model component
{\sc kerrbb{\small 2}}, which we describe below.  This component is a
direct replacement for {\sc diskbb}, returning two fit parameters,
namely the spin and the mass accretion rate (instead of the
temperature and radius of the inner disk).  The four models progress
sequentially in the sense that Model~R1 is the most primitive and
Model~R4 is the most advanced.  This sequence builds toward our
adopted model.  We have chosen to present our results for these
preliminary relativistic models, in addition to those for our adopted
model, because doing so demonstrates that our modeling of the critical
thermal component, and the extreme spin it delivers for Cygnus X-1,
are insensitive to the details of the analysis.

\subsection{Our Adopted Model}

The model we employ is a culmination of Models R1--R4 in the sense
that it is the most advanced and physically realistic model. The
schematic sketch of the X-ray source in Figure~2 illustrates the
various model components and their interplay.  The structure of our
adopted model, naming all the components that comprise it, is
expressed as follows:

\vspace{-3mm}

\begin{eqnarray*}
{\rm CRABCOR*CONST*TBABS~[SIMPLR \otimes KERRBB{\small 2}}\\
{\rm +KERRDISK+~KERRCONV\otimes (IREFLECT \otimes SIMPLC)]}
\end{eqnarray*}

As described in detail below, {\sc simplr} generates the power-law
component using the seed photons supplied by the single thermal
component {\sc kerrbb{\small 2}}, while the reflection component is
likewise generated in turn by {\sc ireflect} acting solely on the
power-law component (i.e., {\sc ireflect} does not act on the thermal
component).  Furthermore, the model fits for a single value of $a_*$,
which appears as the key fit parameter in three model components: {\sc
kerrbb{\small 2}}, {\sc kerrdisk} and {\sc kerrconv}.

We now discuss in turn the model's three principal components --
thermal, power-law and reflected -- and their interrelationships.

{\it Thermal component:} The centerpiece of our adopted model is our
accretion-disk model {\sc kerrbb{\small 2}}, which includes all
relativistic effects, self-irradiation of the disk (``returning
radiation'') and limb darkening \citep{lil+2005}.  The effects of
spectral hardening are incorporated into the basic model {\sc kerrbb}
via a pair of look-up tables for the hardening factor $f$
corresponding to two representative values of the viscosity parameter:
$\alpha=0.01$ and 0.1 \citep{mcc+2006}.  Motivated by observational
data obtained for dwarf novae \citep{sma+1998,sma+1999} and soft X-ray
transients \citep{dub+2001}, and the results of global general
relativistic magnetohydrodynamic (GRMHD) simulations,
\citep{pen+2010}, throughout this work we adopt $\alpha=0.1$ as our
fiducial value; meanwhile, in Section~5.4 we examine the effects on
our results of using $\alpha=0.01$ in place of $\alpha=0.1$.  The
entries in the look-up tables for $f$ were computed using both {\sc
kerrbb} and a second relativistic disk model {\sc bhspec}
\citep{dav+2005,dav+2006b}.  We refer to the model {\sc kerrbb} plus
this table, and the subroutine that reads it, as {\sc kerrbb{\small
2}} \citep{mcc+2006}.  As noted above, the model {\sc kerrbb{\small
2}} has just two fit parameters, namely the black hole spin parameter
$a_*$ and the mass accretion rate $\dot M$ (or equivalently, $a_*$ and
the Eddington-scaled bolometric luminosity, $l \equiv L_{\rm
bol}(a_*,\dot M)/L_{\rm Edd}$).  For the calculations reported in this
paper, we included the effects of limb darkening and returning
radiation.  We set the torque at the inner boundary of the accretion
disk to zero (as appropriate when $D$, $M$ and $i$ are held fixed),
allowed the mass accretion rate to vary freely, and fitted directly
for the spin parameter $a_*$.

{\it Power-law component:} The first term in the sum, {\sc
simplr$\otimes$kerrbb{\small 2}}, models the power-law component and
the observed thermal component in combination.  This dominant part of
the spectrum (see Figure~3) is computed by convolving {\sc
kerrbb{\small 2}}, which describes the seed photon distribution (i.e.,
the thermal component prior to being scattered), with {\sc simplr}.
The convolution model {\sc simplr} is a slightly modified version of
{\sc simpl} \citep{ste+2009b} that is appropriate when including a
separate and additive reflection component \citep{ste+2010b}.  The
parameters of {\sc simplr} (and {\sc simpl}) are the power-law photon
index $\Gamma$ and the scattered fraction, $f_{\rm SC}$, which is the
fraction of the seed photons that are scattered into the power-law
tail.  As used here, {\sc simplr} describes a corona that sends
approximately half the scattered seed photons outward toward the
observer and the remainder downward toward the disk, thereby
generating the reflected component (see below).  Thus, we assume that
the power-law component illuminating the disk is the same as the
component we observe.

{\it Reflected component:} The second and third terms in the sum model
the reprocessed emission from the disk that results from its
illumination by the power-law component.  The model for the
illuminating power-law component itself (the term on the far right) is
{\sc simplc}, which is equivalent to {\sc simplr$\otimes$kerrbb{\small
2}} minus the unscattered thermal component \citep{ste+2010b}.  This
power-law component is acted on by {\sc ireflect}, which is a
convolution reflection model with the same properties as its
widely-used parent, the additive reflection model {\sc pexriv}
\citep{mag+1995}.  Concerning {\sc ireflect}, we (i) free the
reflection scaling factor\footnote{The reflection scaling factor $s$
in {\sc ireflect} (and {\sc pexriv}) is linked to the reflection
constant parameter $x$ in {\sc simplr} via the relation $x$=1+$|s|$
(where $|s|$ is the absolute value of $s$).  In the limiting case of
$x=2$, half of the scattered photons are redirected downward and
illuminate the disc. In the limiting case of $x=1$, none of the
Compton-scattered photons strike the disc \citep{ste+2010b}. For the
preliminary models described in the appendices that use the additive
model {\sc pexriv}, the scaling factor $s$ is fixed at -1.}$s$ while
restricting its range to negative values, thereby computing only the
reflected spectrum, while assuming that the corona is a thin slab that
hugs the disk and emits isotropically; (ii) set the
elemental-abundance switch to unity, which corresponds to solar
abundances, while allowing the iron abundance to be free; (iii) link
the photon index to the value returned by {\sc simplr}; and (iv) fix
the disk temperature at $6.0\times 10^6$ K, the temperature $T_{\rm
in}$ returned by {\sc diskbb} (see Appendix A).  The model {\sc
ireflect$\otimes$simplc} returns a reflected spectrum that contains
sharp absorption features and no emission lines.  To complete the
model of the reflected component, we follow \citet{bre+2006} and
employ the line model {\sc kerrdisk} and the convolution smearing
model {\sc kerrconv}, both of which treat $a_*$ as a free fit
parameter\footnote{We performed tests with a newer version of these
models, {\sc relline} and {\sc relconv}~\citep{dau+2010}, which are
interpolated on a finer grid, and found that our results presented in
Section 4 (and elsewhere) are essentially unchanged.}.  These models
allow the emissivity indices to differ in the inner and outer regions
of the disk.  For simplicity, and because this parameter is unknown
with values that vary widely from application to application, we use
an unbroken emissivity profile with a single index $q$.  We tie
together all the common parameters of {\sc kerrdisk} and {\sc
kerrconv}, including the two principal parameters, namely $a_*$ and
$q$.

The three multiplicative model components are, respectively, (1) {\sc
crabcor}, which corrects for detector effects (see Section~2); (2)
{\sc const}, which reconciles the calibration differences between the
detectors (throughout the paper, we fix the normalization of the {\it
RXTE}/PCU2 detector and float the normalization of the {\it ASCA} GIS
and {\it Chandra} HETG/ACIS detectors); and (3) {\sc
tbabs}\footnote{{\sc tbabs} uses updated values for the photoelectric
cross sections and ISM abundances and is an improved version of the
familiar model {\sc phabs}.  The choice of low-energy absorption model
has a negligible affect on our results apart from the higher column
density ($\sim40$\%) that {\sc tbabs} returns compared to {\sc
phabs}.}, which models low-energy absorption \citep{wil+2000}.
Concerning {\sc tbabs}, throughout the paper we allow $N_{\rm H}$ to
vary because the column density is well-determined by the data and
$N_{\rm H}$ is known to vary by several percent for all three
well-studied supergiant black-hole binaries, namely Cygnus X-1
\citep[][~also see Section~4]{han+2009}, M33 X-7 \citep{liu+2008} and
LMC X-1 \citep{han+2010}.


\section{Results}

The fit results for our adopted model are given in
Table~\ref{table:model_5_results}, and the three fitted spectra
together with their spectral components are shown in Figure~3.  The
fits are all good, with $\chi^2_{\nu}$ ranging from 1.17 to 1.28, and
the results in Table~\ref{table:model_5_results} are in good agreement
with those obtained using Model~R1--R4 (see Appendix~B and Tables
\ref{table:model_1_results}--\ref{table:model_4_results}).  As in the
case of Models~R1--R4, the spin parameter is high with $a_*>0.99$ for
all three spectra.  This is the principal result of this section.  In
the following section, we examine the robustness of this result, and
in Section~6 we fold in the uncertainties in the input parameters $D$,
$M$ and $i$ and arrive at our final lower limit on $a_*$.

Another parameter of great interest is the scattering fraction, which
measures the strength of the Compton component: $f_{\rm SC} =
22.5\pm0.6$\%, $30.5\pm1.2$\% and $30.6\pm0.6$\% for SP1, SP2 and SP3,
respectively. These values are high compared to values characteristic
of the thermal dominant state \citep[$f_{\rm
SC}~\simless~5$\%;][]{ste+2009b}, which is the most favorable state
for the measurement of black hole spin.  These large values of $f_{\rm
SC}$, as well as the uniform value of the power-law index
($\Gamma\sim2.5$), imply that for all three observations the source
was in the steep power-law state \citep{rem+2006}. \citet{ste+2009b}
have shown that one can obtain reliable values of the inner disk
radius (and hence spin) in the SPL state for $f_{\rm
SC}~\simless~25$\% (see their Figure~1).

The luminosity of the disk component, $L/L_{\rm Edd}\approx0.02$, is
quite low and easily meets our data selection criterion, $L/L_{\rm
Edd}<0.3$ \citep{mcc+2006}. Correspondingly, the disk is geometrically
very thin at all radii \citep[the aspect ratio $h/r<0.05$;
see][]{pen+2010,kul+2010}.  At the same time, the luminosity is
sufficiently high that the spectral hardening factor $f$ is
well-determined by the data ($f\approx1.6$).

Interestingly, for the pair of {\it Chandra} observations that are
separated by just two days, the values of the fit parameters are quite
similar (Table~\ref{table:model_5_results}) with the notable exception
of the column density: $ N_{\rm H}=0.768\pm0.024~\rm cm^{-2}$ for SP2
and $ N_{\rm H}=0.687\pm0.010~\rm cm^{-2}$ for SP3.  We tested whether
one can achieve a good fit with a single value of $N_{\rm H}$, by
fitting the {\it Chandra/RXTE} spectra jointly, linking only the
parameter $N_{\rm H}$.  This linking increased the total chi-square by
41, which corresponds to an F-test probability of $3.0\times10^{-9}$.
This low probability implies a significant change in $N_{\rm H}$,
which is likely due to absorption in the wind of the supergiant
companion.  Indeed, one expects such absorption in a wind to be larger
for SP2 at orbital phase 0.24 than for SP3 at phase 0.61, as observed
\citep[Table~1;][]{bau+2000}.


\section{Robustness of Spin Estimates}

In this section we consider a number of factors that might affect our
results and find that none of them is significant.  We consider the
effects of: (1) excluding the Fe~K$\alpha$ line and edge features from
the fits; (2) including HEXTE data and extending the fits to 150~keV;
(3) using {\sc reflionx} to model the reflection component; and (4)
substituting $\alpha=0.01$ for our fiducial value of $\alpha=0.1$ and
varying the metallicity.  As shown below, factors (1) and (2) have
negligible effects on our results, and (3) and (4) have slight upward
effects on $a_*$, which implies that the extreme values reported in
Table~2 are conservative lower limits.  Finally (Section 5.5), we
explore the effects on our results of artificially relaxing the spin
parameter away from its extreme, limiting value by varying the
parameters of {\sc kerrbb{\small 2}}; we find that the input
parameters $D$, $M$, and $i$ have to be driven far from their fiducial
values in order to obtain a spin value as low as $a_*=0.9$.

\subsection{Effect of iron line and edges}

For all three spectra, we refitted the data excluding the energy range
5.0-10.0 keV while omitting the component {\sc kerrdisk}.  This
excised energy range contains the relativistically-broadened
Fe~K$\alpha$ line and edge, as well as a significant residual feature
near 9 keV (Figure~3).  For all three spectra, we find that our
results are essentially unchanged, apart from small shifts in the
parameters of the reflection component.  The results for spectrum SP1
(only) are shown in Table~3 (Case~2) where they are compared to our
standard results (Case~1). Thus, we find that the values of the spin
parameter returned by the fits are completely determined by the
temperature and luminosity of the thermal component and are unaffected
by the presence of the line.  This conclusion is reasonable given that
the line is a minor feature with an equivalent width of
$\approx0.15$~keV, or only $\approx0.015$~keV relative to the
continuum at the peak of the thermal component (see Figure~3).

\subsection{Effect of extending the bandwidth to 150~keV}

In Section~2, we asserted that the coverage of the PCA to 45~keV was
sufficient to adequately constrain the power-law and reflection
components.  We now demonstrate that this is true by refitting the
data for SP1 (only) while including the HEXTE data spanning the energy
range 20 keV to 150 keV (where the source counts are negligible). For
the HEXTE data, we do not a add a systematic error to the count rates.
However, we do correct the detector response to the Crab spectrum of
Toor \& Seward (see Section~2) using the value of the Crab's photon
index as measured using
HEXTE\footnote{\url{http://web.mit.edu/iachec/IACHEC\_2\_talks/IACHEC\_II\_suchy.pdf}}:
$\Gamma=1.93\pm0.003$.  The slope correction is $\Delta \Gamma_{TS} =
-0.17$.  The results obtained using the HEXTE data, which are given in
Table~3 (Case~3), are almost identical to our standard results for SP1
(Case~1).  This is not surprising because the PCA coverage to 45 keV
is more than adequate to determine the slope of the power-law
component, and the reflection component is quite weak and dying
rapidly at 45~keV (Figure~3).

\subsection{Effect of using a different reflection model}

We replace {\sc ireflect$\otimes$simplc+kerrdisk} with {\sc
reflionx}\footnote{The model can be downloaded at \\
\url{http://heasarc.gsfc.nasa.gov/docs/xanadu/xspec/models/reflion.html
}} \citep{ros+2005}, which is widely used in measuring spin via the
Fe~K$\alpha$ line.  The merit of this alternative reflection model is
that it self-consistently calculates the line feature and the
reflection component, whereas {\sc ireflect} models only the
absorption edges.  The downside of {\sc reflionx} is its description
of the seed photon distribution as a simple power-law, which
unphysically diverges at low energies \citep[for further comparison of
the two models, see][]{ste+2010b}. We include relativistic blurring by
convolving {\sc reflionx} with {\sc kerrconv}.  Compared to the
results given in Table~2, the values of the parameters returned by the
fits are generally different, although reasonable, and the fits are
somewhat poorer ($\chi_{\nu}^{2}=$1.35--1.45).  Importantly, the
effects on the spin parameter are very small: The value of $a_*$
increases slightly for SP1 (0.9985 to 0.9999) and is unaffected for
SP2 and SP3, which are at their maximum values.

\subsection{Effect of Varying the Viscosity Parameter and Metallicity}

We refitted the spectra using $\alpha=0.01$ in place of our fiducial
value, $\alpha=0.1$.  Again, the spin parameter for SP1 increases
slightly (0.9985 to 0.9988), while the values for SP2 and SP3 remain
unchanged.  Concerning possible metallicity effects, we do not have
nonsolar-metallicity table models for computing spectral hardening for
such extreme values of spin.  However, for three sources (M33 X-7,
\citeauthor{liu+2008}~\citeyear{liu+2008}; LMC X-1,
\citeauthor{gou+2009}~\citeyear{gou+2009} and A0620--00,
\citeauthor{gou+2010}~\citeyear{gou+2010}), we have found that
substantial changes in metallicity produce very small changes in
$a_*$.  For example, reducing the metallicity from solar to a tenth
solar decreases the spin parameter of a high-spin black hole like LMC
X-1 ($a_*=0.938\pm0.020$) by only $\Delta a_*=0.001$.  The effect of
this same change in metallicity for the slowly-spinning black hole
A0620--00~($a_*=0.135\pm0.029$) is larger, but it is still quite
small: $\Delta a_*=0.014$ .  (The small errors on $a_*$ quoted here
come directly from fitting the X-ray spectra, and they do not take
into account the uncertainties in $D$, $M$ and $i$, which dominate the
error budget).  We note that the super-solar iron abundances implied
by our fits using the reflection models, {\sc ireflect} (Table~2) and
{\sc pexriv} (Appendix B), suggest that the metallicity of the Cygnus
X-1 system is possibly enhanced. Accounting for the supersolar
abundances will result in a slightly increased estimate of $a_*$, so
our conclusions are robust to enhanced metallicity.

\subsection{Relaxing the Spin Parameter}

We now describe three technical exercises that artificially examine
the effects of varying three parameters of {\sc kerrbb{\small 2}},
namely, $a_*$, $D$ and the normalization constant $N_{\rm K}$ (which we
have elsewhere fixed to unity, as is appropriate when $D$, $M$ and $i$
are specified).  Our motivation is to examine the consequences of
relaxing the spin parameter away from the extreme values returned by
the fits (Table~2).

First, for spectrum SP1 only, we leave $N_{\rm K}$ fixed at unity and
allow the distance to vary (keeping $M$ and $i$ fixed at their
fiducial values).  For our parallax distance of $D=1.86$~kpc, we of
course have our standard result, $a_*=0.9985$ (Table~2).  We now
arbitrarily and successively fix the spin parameter at two lower
values, $a_*=0.95$ and $a_*=0.90$, and refit spectrum SP1.  The
best-fit values of $D$ are then substantially greater than our
measured value of 1.86~kpc: $2.43\pm0.03$~kpc and $2.78\pm0.02$~kpc,
respectively. Meanwhile, the corresponding values of reduced
chi-square are respectively 1.34 and 1.40, which are significantly
greater than our standard value of 1.24 (Table~2).  Thus, the best fit
is achieved for the observed and extreme value of spin.

Secondly and alternatively, we obtain similar results by varying
$N_{\rm K}$ while leaving the distance fixed at its fiducial value of
1.86~kpc.  For the same pair of forced values of the spin parameter
given above (0.95 and 0.90), we find for spectrum SP1 that the fitted
values of $N_{\rm K}$ are about half the standard value of unity:
$0.65\pm0.01$ and $0.45\pm0.01$, respectively (where a smaller value
corresponds to a weaker thermal component).  Meanwhile, we find values
of reduced chi-square that are very similar to those given in the
previous example, with values that increase as the spin parameter
decreases.  That is, we again find that the best fit is achieved for
the observed and extreme value of spin.

In a final experiment, we assess the consistency of our spin values
for the three spectra, which can not be adequately judged by
considering the extreme values given in Table~2.  We make this
assessment by artificially setting $N_{\rm K}=0.5$ in our adopted
model and refitting the three spectra.  In this way, we find best-fit
spin values that are in reasonable agreement:
$a_*=0.937_{-0.001}^{+0.001}$ for SP1, $a_*=0.934_{-0.020}^{+0.021}$
for SP2, and $a_*=0.953_{-0.006}^{+0.007}$ for SP3.


\section{Comprehensive Error Analysis}

In all previous work on measuring spin, we have found that the
statistical uncertainty in the estimated spin parameter due to the
X-ray data analysis is small compared to that due to observational
uncertainties in $D$, $M$ and $i$.  Based on our GRMHD simulations of
thin disks \citep{sha+2008,pen+2010}, we have likewise found that the
uncertainties in these three input parameters dominate over the errors
resulting from our use of the analytic Novikov-Thorne model, which
assumes a razor-thin disk.  In the case of Cygnus X-1, these model
errors are especially small because of the extreme spin of the black
hole and the low luminosity of the disk.  Spin estimates obtained by
fitting mock spectra of simulated GRMHD disks indicate that for an
inclination $i=30$~deg, which is very near the inclination of Cygnus
X-1 ($i=27.1$~deg), the Novikov-Thorne thin-disk model overestimates
the spin parameter by only $\Delta{a_*} = 0.007$ and $0.005$ for spins
of 0.90 and 0.98, respectively \citep[see Table~1 in][]{kul+2010}.
Furthermore, these errors are significantly overestimated because they
were computed for disks that are far more luminous ($L/L_{\rm
Edd}\sim0.5$), and hence thicker, than that of Cygnus~X-1 ($L/L_{\rm
Edd}\approx0.02$)\footnote{It is computationally very challenging to
simulate thinner disks.}. In our error analysis, we neglect this small
model error.

With the exception of our recent spin measurement of XTE J1550--564
\citep{ste+2010b}, in earlier work we have neglected the uncertainty
in the luminosity due to the $\sim10$\% uncertainty in the flux of the
Crab \citep{too+1974}, an error that uniformly shifts all of our spin
measurements either up or down.  For the spin of Cygnus X-1, the
effect of the uncertainty in the absolute flux calibration is very
comparable to the 6\% uncertainty in $D$ (which is equivalent to a
12\% uncertainty in the measurement of flux).  We have therefore
included in our error budget the 10\% uncertainty in flux (which we
approximate as an uncertainty in the distance of 0.1 kpc) by simply
combining the distance and flux-calibration errors in quadrature,
thereby inflating the actual~0.120 kpc distance uncertainty to
0.156~kpc.  Thus, the final error we report for $a_*$ includes the
uncertainties in $D$, $M$, $i$ and the uncertainty in the absolute
flux calibration.  Taken together, these four sources of uncertainty
totally dominate the error budget.

In order to determine the error in $a_*$ due to the combined
uncertainties in $D$, $M$ and $i$, we performed Monte Carlo
simulations using the Odyssey computing cluster at Harvard University.
The latter two parameters are not independent.  They are related
through the expression for the mass function: $f(M)\equiv
{M^3}$sin$^3i/{(M+M_{\rm opt})^2}$ = $0.263\pm0.004~M_{\odot}$, where
$M_{\rm opt}=19.16\pm1.90~M_{\sun}$ is the mass of the secondary star
and the value of the mass function was evaluated using a K-velocity of
$76.79\pm0.41~{\rm km~s^{-1}}$ and an orbital period of $P=5.599829$
days \citep{oro+2010}.  

In the analysis, we assumed that the value of the mass function, the
inclination, and the mass of the secondary are normally and
independently distributed, and we computed the mass of the black hole
using the values of these quantities, which are given above.  We
conservatively fixed the viscosity parameter at its baseline value,
$\alpha=0.1$ (using $\alpha=0.01$ increases $a_*$; see Section~5.4).
Specifically, we (i) generated 9000 parameter sets for $D$, $i$,
$M_{\rm opt}$, and $f(M)$; (ii) solved for $M$ for a given triplet of
values of $i$, $M_{\rm opt}$ and $f(M)$; (iii) computed for each set
the look-up table for the spectral hardening factor $f$ using the
model {\sc bhspec}; and (iv) obtained $a_*$ by fitting our adopted
model to the spectra.  The final histogram distributions for our three
spectra are shown in Figure~4.  Consulting these histograms, we see
that the spin estimate is lowest for SP1, and we conservatively base
our final result on this spectrum, {\it thereby concluding that
$a_*>0.95$ at the $3\sigma$ level of confidence (see Figure~4 and the
bottom line of Table~2).}

We note the following two caveats: First, the power-law component is
strong relative to the thermal component, which decreases the
reliability of the continuum-fitting method
\citep{ste+2010a,ste+2010b}.  Second, as we discuss in Section~7.4,
the continuum-fitting method assumes that the spin vector of the black
hole is aligned with the orbital angular momentum vector.


\section{\sc discussion}

Our wide-ranging discussion covers seven topics: (1) We first confront
the challenge posed by a grossly discrepant spin result that was
obtained using the Fe-line method (while noting a concordant Fe-line
result that appeared very recently), and (2) a second discrepant
result obtained using a QPO method.  (3) We next consider a disfavored
dynamical model that implies a less extreme value of spin.  (4) The
fundamental assumption of the continuum-fitting method, namely the
alignment of the spin and orbital vectors, is then considered.  (5)
The extreme spin of the black hole is reconciled with the low
temperature of its accretion disk.  (6) We find no evidence in our
{\it Chandra} HETG spectra for warm absorbing gas, and we show that,
if it were present, its effects on our spin estimates would be
negligible.  (7) Finally, we discuss some consequences of the extreme
spin of Cygnus X-1, and we argue that this spin is chiefly natal in
origin.

\subsection{Measurement of Spin via the Fe~K$\alpha$ Method}

Based on an analysis of the Fe~K$\alpha$ line profile,
\citet{mil+2005} found marginal evidence for high spin using the same
{\it ASCA} spectrum of Cygnus X-1 that we use.  Subsequently, however,
\citet{mil+2009} reported a spin of $a_*=0.05\pm0.01$ based on an
analysis that combines Fe~K$\alpha$/reflection models and continuum
models, including {\sc kerrbb}, a relativistic disk model similar to
the one we use.  That is, the authors jointly applied the
continuum-fitting and Fe~K$\alpha$ methods and fitted simultaneously
for a single value of $a_*$.  The data analyzed in this case were a
pair of 0.5~ks {\it XMM-Newton/}/EPIC-pn spectra obtained in the
``burst'' timing mode.

While it is not possible for us to account in detail for the gross
difference between the near-zero spin reported by \citet{mil+2009} and
our near-extreme value, we note the following: First, the round values
of $D$, $M$, and $i$ that~\citeauthor{mil+2009}~used as input to {\sc
kerrbb} differ significantly from ours, and these differences all
serve to drive the spin down; e.g., using their values of these
parameters and our spectrum SP1, we find $a_*=0.74\pm0.01$.  Secondly,
and in addition,~\citeauthor{mil+2009}~fitted for the {\sc kerrbb}
normalization constant, obtaining $N_{\rm K}=0.31$, whereas the
standard procedure, which we follow, is to set $N_{\rm K}$ to unity
when $D$, $M$, and $i$ are fixed.  Fitting SP1 using $N_{\rm K}=0.31$
(and the values of $D$, $M$, and $i$ adopted by
~\citeauthor{mil+2009}), we find that the spin drops from $a_*=0.78$
to the retrograde value $a_*=-0.53$ (and the fit is poor,
$\chi_{\nu}^{2}=2.01$.)

The near-zero spin result of \citet{mil+2009} is further called into
question because no data above 10~keV were used.  This lack of high
energy coverage, in the presence of a strong Compton component,
seriously compromises results obtained using the continuum-fitting
method, as we stress at the outset of Section~2 (and as can be deduced
by an examination of Figure~3).  Likewise, results obtained using the
Fe~K$\alpha$/reflection method are compromised by the failure to
observe the Compton reflection hump around 20--30 keV \citep[e.g.,
see][]{lar+2008}.

Very recently, after posting our paper on the astro-ph archive and
while revising it for resubmission, a paper appeared reporting another
estimate of the spin of Cygnus X-1 via an analysis of the Fe~K$\alpha$
profile \citep{dur+2011}.  Assuming that the emissivity profile of the
disk can be described by a single power-law with index 3.0, these
authors conclude that ``the black hole is close to rotating
maximally," which is in agreement with our result.  We note that
\citeauthor{dur+2011}\ do not discuss \citet{mil+2005,mil+2009} or
mention the near-zero spin result reported in the latter paper.

\subsection{Measurement of Spin via a QPO Model}

Based on an analysis of low-frequency (0.01--25~Hz) quasi-periodic
oscillations (QPOs), \citet{axe+2005} obtained a spin for Cygnus X-1
of $a_*=0.49\pm0.01$ for $M=8M_{\sun}$.  Their result is based on the
relativistic precession model of \citet{ste+1999}.  In this model, the
QPO is produced as a result of emission from an orbiting bright spot
that is undergoing relativistic nodal and periastron precessions in a
slightly tilted and eccentric orbit.  The spin parameter is predicted
to vary with mass as $a_*\propto M^{-1/5}$ \citep[see Eqn. 4
in~][]{axe+2005}, and therefore the corrected value of spin is
$a_*=0.43$ for our adopted black hole mass $M=14.8~M_{\sun}$.  The
large discrepancy between this moderate value of spin and the extreme
value we find may be a consequence of a fundamental assumption of
their model, namely, that the black hole is rotating slowly ($a_*\ll
1$).  On the other hand, the precession model, with its assumption of
geodesic motion, may not apply in this instance.

\subsection{An Alternative Dynamical Model}

In \citet{oro+2010}, results are presented for four dynamical models,
Models {$\cal A-D$}.  We disregard Models {$\cal A$} and {$\cal B$},
which assume a circular orbit, because these models give poor fits
compared to the eccentric orbit models ($\Delta \chi^{2} >50$) and
because there is clear evidence that the orbit is eccentric.
Throughout this paper, we have used $M=14.8\pm1.0~M_{\sun}$ and
$i=27.1\pm0.8$ deg from Model $\cal D$, an asynchronous model with a
rotational frequency for the O-star that is 40\% greater than the
orbital frequency.  As an alternative to Model {$\cal D$}, we now
consider Model {$\cal C$}, which assumes synchronous rotation.  Model
{$\cal C$} gives a poorer fit to the data ($\Delta \chi^2 \approx
13$), and it results in disharmony between the light-curve and
velocity data on the one hand, and the radius and rotational velocity
of the O-star on the other \citep[see Table~1 in][]{oro+2010}.
Because of this disharmony, the uncertainties in $M$ and $i$ for Model
{$\cal C$} are significantly larger than those for our favored model,
although the central values of these parameters differ only modestly:
$M=15.8\pm1.8~M_{\sun}$ and $i=28.5\pm2.2$ deg.  Using these values
for Model {$\cal C$} and spectrum SP1 (which gives the lowest value of
spin), we repeated our error analysis (Section 6) and obtained the
following 3$\sigma$ lower limit on the spin parameter: $a_*>0.92$.

\subsection{Alignment of Spin and Orbital Angular Momentum}

There remains one uncertainty that calls into question the reliability
of the continuum-fitting method, namely, whether the inner X-ray
emitting portion of the disk (which will align with the black hole's
spin axis) is aligned with the binary orbital plane.  For a discussion
of this question, see Section~2.2 in \citet{lix+2009}.  As an
extension of this discussion, we note the following: First, recent
population synthesis studies predict that the majority of systems will
have rather small ($\lesssim10$~deg) misalignment angles
\citep{fra+2010}.  Second, in the case of Cygnus X-1, there is reason
to believe that the misalignment angle is especially small because of
the binary system's low peculiar velocity, which indicates that the
system did not experience a large ``kick'' when the black hole formed
\citep{mir+2003,rei+2010}. As demonstrated in Figure 5, even if there
exists a misalignment angle as large as (e.g.) 16 degrees, the spin
value is still $>$0.95.

\subsection{Low Disk Temperature Compatible with Extreme Spin}

One might expect that the accretion disk of a fast-spinning black hole
like Cygnus~X-1 would be hot ($T_{\rm in}>$1~keV) because its inner
edge is relatively close to the event horizon and deep in the
gravitational potential well.  Two principal factors are responsible
for the depressed disk temperature ($T_{\rm in} \approx 0.5$~keV;
Table~\ref{table:model_c_results}).  The first of these is the low
rate of mass accretion through the disk, which is manifested by its
low luminosity, only $\approx2$\% of Eddington
(Table~\ref{table:model_5_results}).  The second is that relativistic
effects (e.g., beaming and light bending) are muted for low disk
inclinations: Using our fiducial values of $D$, $M$, and $i$, and fixed
values of the fit parameters taken from Table~2, we used {\sc
kerrbb{\small 2}} to simulate a pair of spectra, one for the
inclination of Cygnus X-1, $i=27.1$~deg, and the other for
$i=66.0$~deg \citep[which is the inclination of
GRS~1915+105;][]{mcc+2006}.  The luminosities of the two spectra are
the same, $L/L_{\rm Edd}\approx0.018$, while the disk temperatures are
$T_{\rm in}=0.45$~keV and 0.86~keV, respectively.  Thus, in this
comparison, the temperature of the low-inclination disk is depressed
by nearly a factor of two because the relativistic effects are weak.

\subsection{Effects of a Warm Absorber}

Although careful modeling of warm absorbers (WAs) is usually crucial
in determining the spins of supermassive black holes via the
Fe~K$\alpha$ method \citep[e.g.,][]{bre+2006}, we find that the
effects of WAs are unimportant in estimating the spin of Cygnus X-1
via the continuum-fitting method.  We examined all of the available
soft-state {\it Chandra} HETG spectra of Cygnus X-1 (ObsIDs 2741,
2742, 2743, 12313/SP2, and 12314/SP3) at $E<1$~keV, and we find no
evidence for the blend of absorption lines near 0.76~keV (due, e.g.,
to Fe~{\sc X}--Fe~{\sc XV}, O~{\sc VII} and O~{\sc VIII}), which is a
telltale signature of a WA.  (The lines WAs produce above 1 keV are
mostly discrete and weaker and therefore have an accordingly much
smaller effect on the continuum shape and our spin estimates.)

Our preliminary analysis of the low-energy portion ($E<0.8$ keV) of a
50 ks {\it hard-state Chandra} HETG spectrum \citep{han+2009} did
reveal the presence of a single WA.  Its two most relevant parameters
are its column density, $N_{\rm H}=7.2 \times 10^{20}$~cm$^{-2}$, and
ionization parameter, $\xi=44$.  (Its turbulent broadening and
redshift are respectively $\rm v=47.6$~km~s$^{-1}$ and $z=-0.00086$;
we assume solar abundances.)  This tenuous WA does not affect the
continuum shape, but it does produce line blends, the strongest of
which is centered at 0.76~keV.  Although such a WA is not present in
soft-state spectra, we nevertheless tested its effects by reanalyzing
SP1, SP2 and SP3 by introducing the additional model component {\sc
WARMABS}, fixing its parameters to the values given above; we obtained
values of spin that are essentially identical to those given in
Table~2. In one further test, we likewise reanalyzed our three spectra
including a second thicker and hotter WA that generates a complex of
absorption lines between 0.9 and 1.0 keV ($N_{\rm H}=1.0 \times
10^{21}$~cm$^{-2}$; $\xi=300$; $\rm v=100$~km~s$^{-1}$; $z=-0.00086$).
Again, the effects on our spin estimates are negligible.

\subsection{Consequences of Extreme Spin and its Origin}

The spin of Cygnus X-1 is extraordinarily high, placing it in the
company of the microquasar GRS 1915+105, estimated using the
continuum-fitting method \citep{mcc+2006,blu+2009}, and the
supermassive black hole in MCG-6-30-15, estimated using the Fe-line
method \citep{bre+2006}. The spins of both of these black holes are
reported to exceed $a_*=0.98$. As \citet{pen+1969} first demonstrated,
the enormous ``flywheel'' energy of a fast-spinning black hole can in
principle be tapped.  For Cygnus X-1, the potentially tappable energy
is $>$2.8~\msun$c^2 =5.0 \times 10^{54}$~ergs \citep{chr+1971}; in
comparison, the energy radiated by the Sun over its entire
$\sim10$~billion-year lifetime is $\lesssim0.001$~\msun$c^2$.  It has
been widely suggested that spindown energy powers the relativistic
jets observed for at least some quasars and microquasars
\citep{bla+1977}, such as GRS~1915+105 \citep{mir+1994}.

As the spin $a_*$ of a black hole approaches unity, the radius of its
ISCO approaches the radius of the event horizon $R_{\rm EH}$
\citep{bar+1972}. For Cygnus X-1, with spin $a_* > 0.95$ and
$M=14.8$~\msun, $R_{\rm ISCO}<42$~km while $R_{\rm EH}<29$~km.  The
Keplerian velocity of the gas at the ISCO is approximately half the
speed of light, and its orbital frequency is $>598$~Hz.  Meanwhile,
the frequency of rotation of the black hole itself, which is the spin
frequency of space-time at its horizon, is $>790$~Hz (the maximal spin
frequency for $a_*=1$ is 1091 Hz).

What is the origin of the spin of Cygnus X-1?  Was the black hole born
with its present spin, or was it torqued up gradually by the gas it
has accreted over its lifetime?  To achieve a spin of $a_*>0.95$ via
disk accretion, an initially nonspinning black hole must accrete
$>7.3$~\msun~from its donor \citep{bar+1970,kin+1999} in becoming the
$M=14.8$~\msun~that we observe today. However, to transfer this much
mass even at the maximum (Eddington-limited) accretion rate requires
$>31$ million years\footnote{The corresponding estimates of spin-up
time we reported for both M33~X-7 \citep{liu+2008,liu+2010} and
LMC~X-1~\citep{gou+2009} are incorrect because for the efficiency
$\eta$ (of converting mass to radiant energy) we used a constant
value, $\eta=1$.  However, $\eta$ gradually increases as the black
hole spins up, starting at $a_*=0$ with $\eta=0.06$ to $\eta=0.13$
($a_*=0.84$) for M33~X-7 and $\eta=0.17$ ($a_*=0.92$) for LMC~X-1.
For M33~X-7 and LMC~X-1, the correct limits on the spin-up timescales
are $>17$ and $>25$ million years, respectively, while the respective
ages of the systems are $\lesssim3$ and $\lesssim5$ million years.},
whereas the age of the binary system is between 4.8 and 7.6 million
years\footnote{A 7.6 million-year-old system accreting at the maximum
rate would only achieve a spin of 0.59.}  \citep{won+2011}. (Even for
$a_*>0.92$ obtained for our less probable model, the accretion
timescale is $>$ 25 million years.)  Thus, it appears that the spin of
Cygnus X-1 must be chiefly natal \citep[also,
see][]{axe+2011,won+2011}, although possibly the high spin could be
achieved during a short-lived evolutionary phase of hypercritical mass
accretion \citep{mor+2011}.

\section{Conclusion}

Based on our favored dynamical model, we find an extreme value of spin
for the black hole primary in Cygnus X-1: $a_*>0.95$ at the 3$\sigma$
level of confidence.  For a less probable (synchronous) dynamical
model, the spin is still high: $a_*>0.92$ (3$\sigma$).  For both of
these strong limits, we include the customary uncertainties in the
input parameters $D$ (6\%), $M$ (7\%) and $i$~($\pm0.8$ deg), and we
also include the uncertainty in the calibration of the absolute flux
via the Crab (10\%).  These four sources of uncertainty totally
dominate the error budget.

Our measurement of spin is determined solely by the properties of the
thermal component and is unaffected by the presence of the relatively
faint Fe~K$\alpha$ line.  Nevertheless, we have modeled this
relativistically-broadened line feature carefully in order to achieve
good fits over the full range of energies we consider, which in our
analyses variously extends from 0.5~keV to 150~keV.  The extreme spin
we find for this black hole is based on an analysis of three broadband
spectra that are each capable of constraining the soft thermal
component, the hard Compton component, and the reflected component. By
considering several different models and performing a number of tests
on our results, we have demonstrated that the extreme spin we find is
insensitive to the details of our analysis.

\acknowledgements

We thank an anonymous referee for very helpful and constructive
comments. We are grateful to Director H. Tanananbaum and Project
Scientist T. Strohmayer for granting us respectively {\it Chandra} and
{\it RXTE} observing time.  We thank H. Marshall, M. Nowak and
N. Schulz for help in planning the {\it Chandra} observations, and
M. Hanke, M. Nowak and J. Wilms for discussions on X-ray data
analysis.  The research has made use of data obtained from the High
Energy Astrophysics Science Archive Research Center (HEASARC) at
NASA/Goddard Space Flight Center.  LG thanks the Harvard FAS Sciences
Division Research Computing Group for their technical support on the
Odyssey cluster. JEM acknowledges support from NASA grants DD0-11049X,
DD1-12054X and NNX11AD08G, and the Smithsonian Endowment Funds.


\clearpage

 \begin{figure} [f!]
    \begin{center}
      \includegraphics[angle = 0, trim =0cm 0cm 0cm 0cm,width=1.0\textwidth]{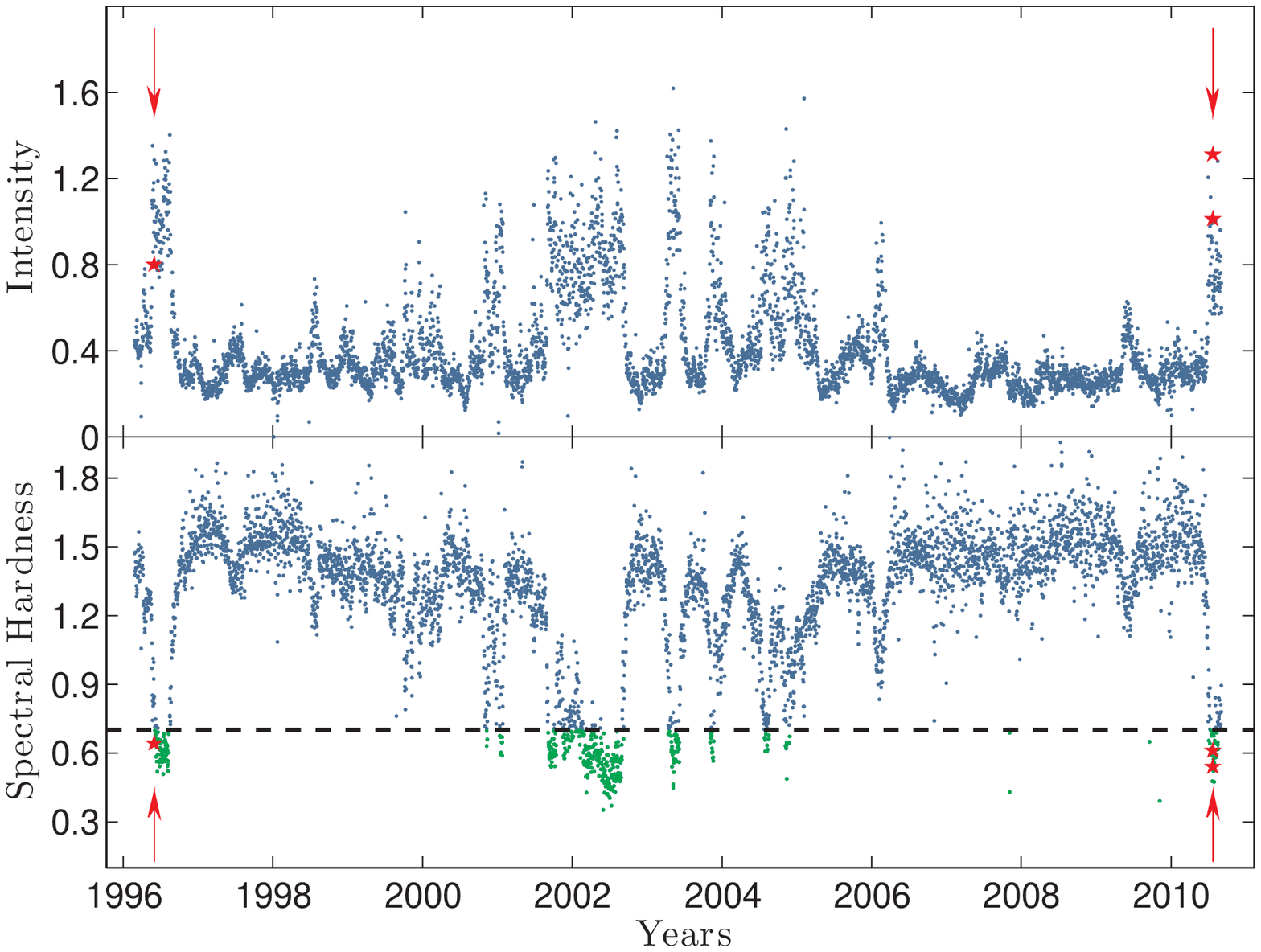}
       \end{center}
       {\bf Figure 1.}  Fourteen-year record for Cygnus~X-1 of X-ray
intensity relative to the Crab nebula ({\bf top}) and spectral
hardness SH ({\bf bottom}).  The hardness is defined as the ratio of
counts detected in a hard X-ray band (5--12~keV) to those detected in
a soft (1.5--5~keV) band.  We consider data suitable for the
measurement of spin only when the spectral hardness is below the
dashed line (SH~$<~0.7$), which is an empirical choice.  Shown plotted
as red stars are the intensity and hardness of the source for each of
the three selected observations, SP1 in 1996, and SP2 and SP3 in 2010.
While a useful diagnostic for the purposes of data selection, the
{\it RXTE}/ASM sky survey data shown here are unsuitable for the
measurement of spin.
 \end{figure}

 \begin{figure} [ftbp]
    \begin{center}
      \includegraphics[angle = 0, trim =0cm 0cm 0cm 0cm,width=1.0\textwidth]{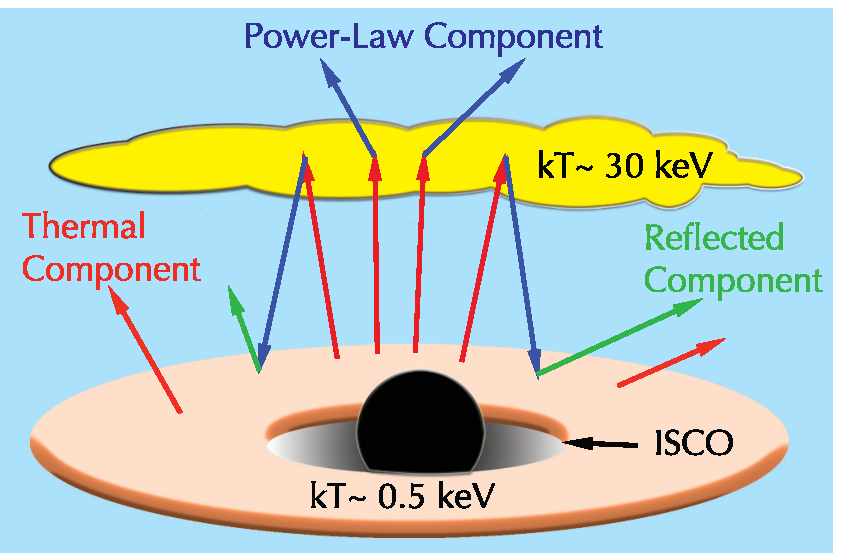}
       \end{center}
 {\bf Figure 2.} Schematic sketch of the X-ray source (adapted from a
sketch provided by R.~Reis).  The accretion disk (pink) is truncated
at the ISCO, leaving a dark gap between the disk's inner edge and the
black hole's event horizon (black).  Shown hovering above the
optically-thick disk is its tenuous scattering corona (yellow).  As
indicated by the arrows, the disk supplies the thermal component of
emission, which is Compton scattered into a power-law component by hot
electrons in the corona.  Approximately half of this latter component
illuminates the disk, thereby generating the reflected component.
\end{figure}

 \begin{figure} [f!]
    \begin{center}
      \includegraphics[angle = 0, trim =0cm 0cm 0cm 0cm,width=1.0\textwidth]{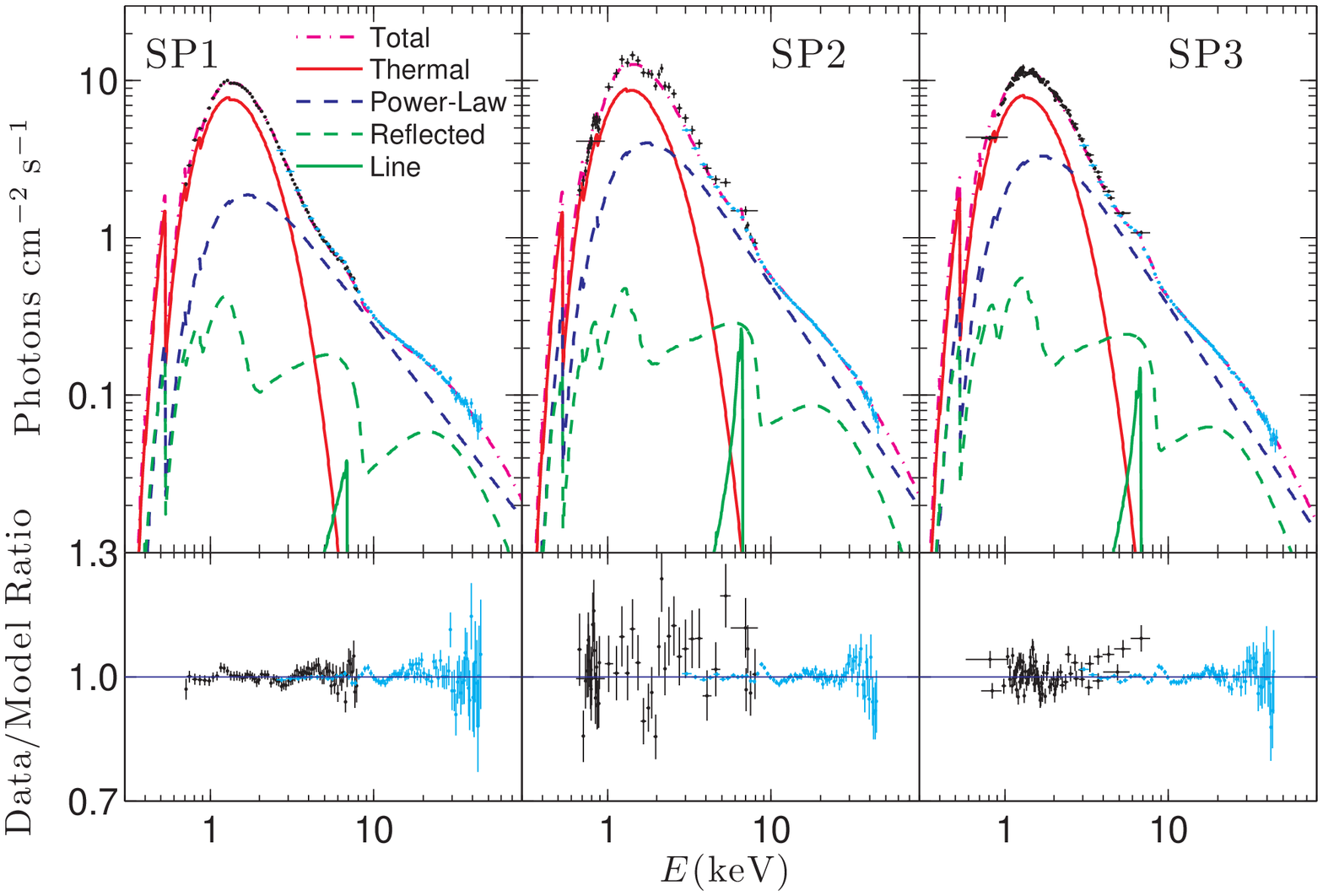}
       \end{center}
{\bf Figure 3.} ({\bf Top}) The upper envelope in each of these
spectra shows the data ({\it RXTE} in blue, and {\it ASCA} or {\it
Chandra} in black) and the best-fit total relativistic model for the
case of our adopted model.  Each total model spectrum is shown
decomposed into thermal and power-law components, and a reflection
component, which is comprised of a continuum component plus the
F~K$\alpha$ line feature.  (The color assignments correspond to those
used in Figure~2.)  The low-energy X-ray absorption component is evident
at energies $\simless1$~keV.  Note in all three spectra the dominance
at low energies of the key thermal component.  ({\bf Bottom}) Ratio of
the data to the model showing deviations between the two.
\end{figure}

 \begin{figure} [f!]
    \begin{center}
      \includegraphics[angle = 0, trim =0cm 0cm 0cm 0cm,width=1.0\textwidth]{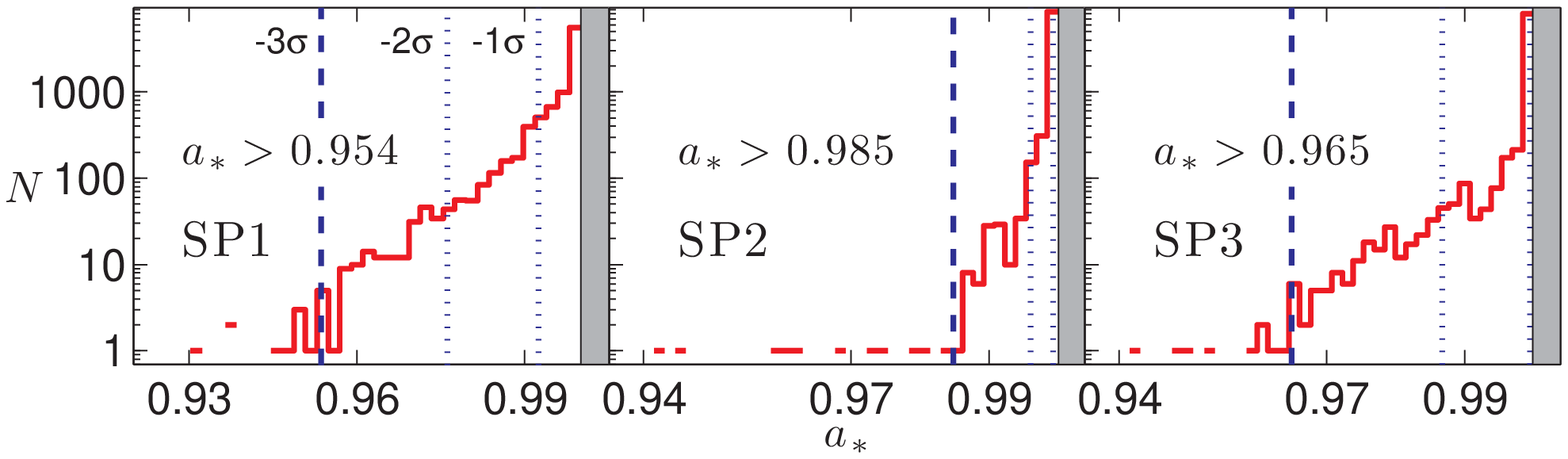}
       \end{center}
       {\bf Figure 4.} Histograms of $a_*$ for 9000 parameter sets
       (per spectrum) resulting from the Monte Carlo analysis.  The
       lower limits given are at the $3\sigma$ (99.7\%) level of
       confidence.
 \end{figure}

 \begin{figure} [f!]
    \begin{center}
      \includegraphics[angle = 0, trim =0cm 0cm 0cm 0cm,width=1.0\textwidth]{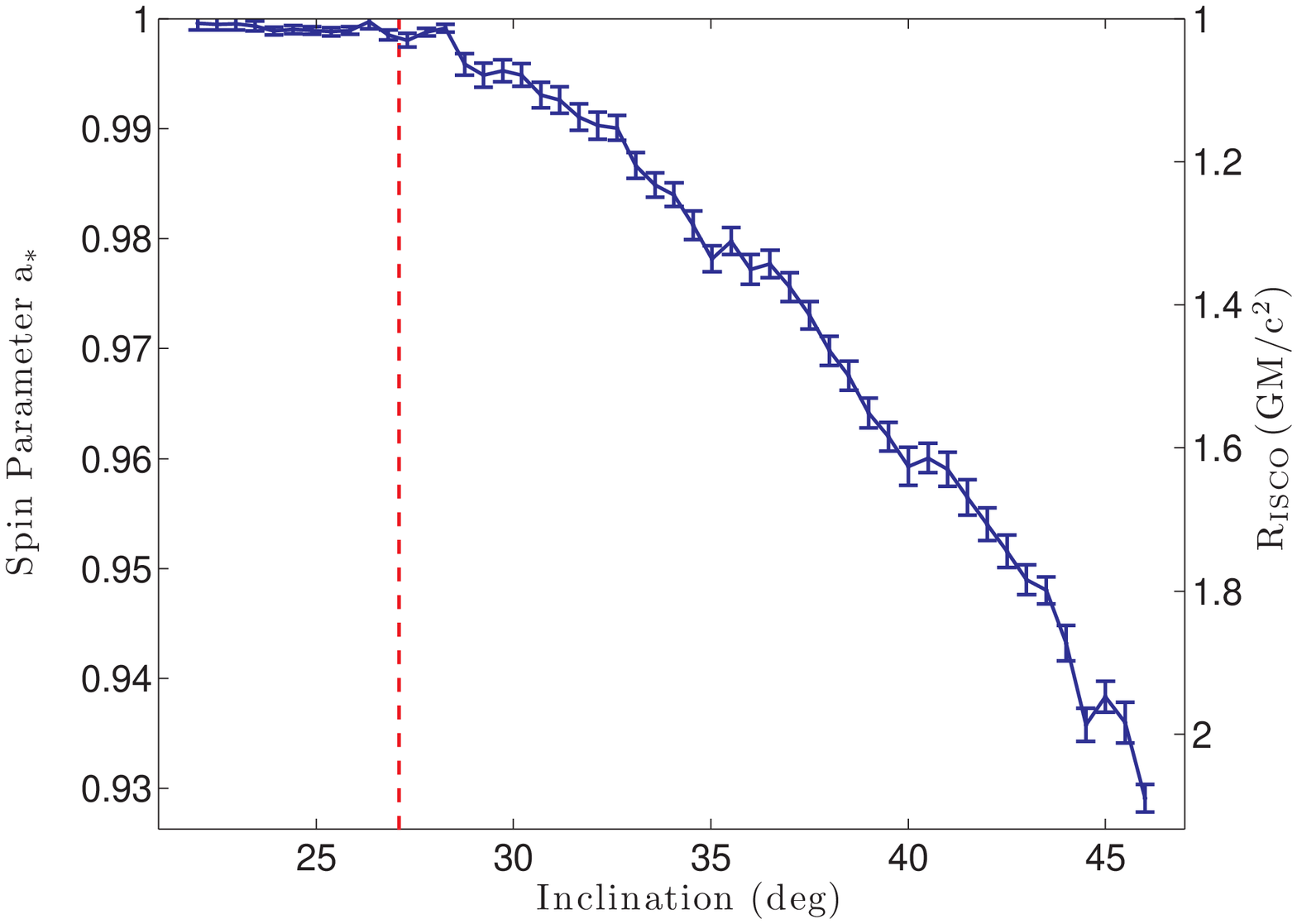}
       \end{center}
{\bf Figure 5.} The effect on the spin parameter of varying the
inclination angle $i$ (for SP1 only) for fixed values of our adopted
parameters: $D=1.86$~kpc and $M=14.8~M_{\odot}$.  The best-fit value
of inclination ($i=27.1$~deg) is indicated by the dashed line.  While
the spin parameter in this figure varies modestly from 0.93 to unity,
the ISCO radius -- the quantity that we actually measure -- can be
seen to change by a factor of 2 (Bardeen et al. 1972).  As an extreme
example, as $a_*$ increases from 0.98 to 0.99, the fractional change
in the ISCO radius is 10 times the fractional change in the spin
parameter.  This "saturation" of the spin parameter near unity is the
reason that the statistical uncertainties given in Table~2 are so
small.
\end{figure} 

\clearpage
\newpage

\begin{deluxetable}{ccccccccc}
\tablewidth{0pt}
\tablecaption{Journal of the observations that yielded spectra SP1,
SP2 and SP3\tablenotemark{a}}
\tablehead{\colhead{No.}& \colhead{Mission} & \colhead{Detector}
  &\colhead{$E_1$-$E_2$ (keV)}
  &\colhead{UT} &
  \colhead{$T_{\rm exp}$ (sec)} & \colhead{I(Crab)} & \colhead{SH} & \colhead{$\phi$}
}

\startdata
SP1 &  {\it ASCA~\&~RXTE}&GIS~\&~PCA &0.7-8.0
\&2.5-45.0 & 1996-05-30 06:43:16~\&~07:51:29 &  2547~\&~2240&0.80 &0.64 &0.74   \\
SP2 & {\it Chandra~\&~RXTE} &ACIS(TE)/HETG~\&~PCA
 &0.5-10.0~\&~2.5-45.0 &2010-07-22 16:21:22~\&~17:04:01   &  2146~\&~3808&1.31&0.54 &0.24   \\
SP3 & {\it Chandra~\&~RXTE}
&ACIS(CC)/HETG~\&~PCA
&0.5-10.0~\&~2.5-45.0 &2010-07-24 17:21:43~\&~17:43:00 &
900~\&~3904&1.01&0.61 &~~0.61
\enddata
\tablenotetext{a}{For each pair of missions listed, in columns 3--6 we
give respectively the following information: names of the detectors
employed, bandwidths used in the analysis, UT start times of the
observations and effective exposure times.  The source intensity and
spectral hardness (SH), which are defined and plotted in Figure~1, are
given in the two columns that follow.  The final column gives the
orbital phase of the binary system defined as the time of supergiant
superior conjunction (black hole beyond O-star), which occurred on
heliocentric Julian Day 2441874.71 \citep{bro+1999}.}

\end{deluxetable}

\begin{deluxetable}{cccccc}
\tablewidth{0pt}
\tablecaption{Fit results for our adopted model\tablenotemark{a}}
\tablehead{\colhead{Number}& \colhead{Model} & \colhead{Parameter}
  &\colhead{SP1}
  &\colhead{SP2} &
  \colhead{SP3}
}

\startdata
1& {\sc kerrbb2  } & $a_* $  & $0.9985_{-0.0008}^{+0.0005} $\tablenotemark{b}& $ 0.9999_{-0.0029}^{+0} $\tablenotemark{b} & $ 0.9999_{-0.0001}^{+0} $\tablenotemark{b} \\
2& {\sc kerrbb2   } & $\dot{M} $  & $ 0.115  \pm 0.004 $& $ 0.139 \pm 0.029 $ & $ 0.122 \pm0.011  $\\
3& const  & -- & $1.000 \pm 0.002$& $1.031 \pm 0.013$ & $0.971 \pm 0.004$ \\
4& {\sc tbabs}  & $N_{\rm H}$ & $0.705 \pm 0.006$& $0.768 \pm 0.024$ & $0.687 \pm 0.010$ \\
5& {\sc simplr  } & $\Gamma $  & $ 2.282 \pm 0.010  $& $ 2.499 \pm 0.012 $ & $ 2.549 \pm 0.011 $\\
6& {\sc simplr  } & $f_{\rm SC} $  & $ 0.225  \pm 0.006 $& $ 0.305 \pm 0.012 $ & $ 0.306 \pm 0.006 $\\
7& {\sc kerrdisk  } & $E_{\rm L} $  & $ 6.56  \pm 0.09 $& $ 6.44  \pm 0.05  $ & $ 6.49 \pm0.04  $\\
8& {\sc kerrdisk } & $q $  & $ 2.82  \pm 0.02 $& $  2.28\pm 0.06 $ & $ 2.29 \pm 0.07 $\\
9& {\sc kerrdisk } & $N_{\rm L} $  & $ 0.015 \pm 0.001 $& $  0.032\pm 0.002 $ & $ 0.024 \pm 0.001 $\\
10& {\sc kerrdisk } & $\rm EW$ & 0.154 & 0.158 & 0.159 \\
11& {\sc ireflect } & $\rm X_{Fe}$  & $ 5.34 \pm 0.15 $& $ 2.97 \pm 0.12 $ & $ 3.63 \pm 0.14 $\\
12& {\sc ireflect } & $ s$  & $ 0.98 \pm 0.04 $& $ 0.80 \pm 0.03 $ & $ 0.93 \pm 0.03 $\\
13& {\sc ireflect } & $\xi $  & $ 153.1 \pm 15.7 $& $ 71.2 \pm 10.0 $ & $ 62.0 \pm 8.4 $\\
\hline
14&  &$\chi^2_{\nu}$  & 1.24(561/454)&  1.28(371/290) &  1.18(723/614)\\
15&  &  $f$ &1.610  &1.612  &1.621  \\
16& &$L/L_{\rm Edd}$ & 0.018& 0.026 &~~0.023\\
\hline
17& Adopted \tablenotemark{c}  & $a_* $  & $0.9985_{-0.0148}^{+0.0005} $& $ 0.9999_{-0.0050}^{+0} $& $ 0.9999_{-0.0116}^{+0} $
\enddata

\tablenotetext{a}{For the model components given, the parameters from
top to bottom are: (1) spin parameter; (2) mass accretion rate in
units of $10^{18}$ g~s$^{-1}$; (3) detector normalization constant
relative to {\it RXTE} PCU2; (4) hydrogen column density in units of
cm$^{-2}$; (5) photon power-law index $\Gamma$; (6) scattering
fraction $f_{\rm SC}$; (7) central line energy in keV; (8) emissivity
index $q$; (9) line flux in units of photons~cm$^{-2}$~s$^{-1}$; (10)
equivalent width of line in keV; (11) iron abundance relative to
solar; (12) reflection scaling factor $s$; (13) ionization parameter
$\xi$; (14) Reduced chi-square, total chi-square and degrees of
freedom, respectively; (15) spectral hardening factor $f$; and (16)
Eddington-scaled disk luminosity, where $L_{\rm Edd}\approx
1.9\times10^{39}~{\rm erg~s^{-1}}$ for Cygnus X-1. Unless otherwise
indicated, the uncertainties quoted here and throughout the paper are
at the $1\sigma$ level of confidence.}

\tablenotetext{b}{The physical maximum value of the spin parameter
$a_*$ is unity and for the XSPEC model {\sc kerrbb2} it is 0.9999.
The very small errors quoted here are purely the uncertainties due to
counting statistics, which result from fitting our adopted model to the
X-ray data.}

\tablenotetext{c}{Final adopted values for the spin parameter and
  their uncertainties. The 1$\sigma$ uncertainties are calculated
  based on the 3$\sigma$ lower limits on $a_*$ shown in Figure~4.
  These results fold in the uncertainties in $D$, $M$, $i$ and the
  absolute flux calibration via our Monte Carlo analysis (see
  Section~6).}

\label{table:model_5_results}
\end{deluxetable}

\begin{deluxetable}{cccccc}
\tablewidth{0pt}
\tablecaption{Effects of excluding the Fe K$\alpha$ line and increasing bandwidth
(SP1 only)\tablenotemark{a}}
\tablehead{\colhead{Number}& \colhead{Model} & \colhead{Parameter}
  &\colhead{Case~1}
  &\colhead{Case~2} &
  \colhead{Case~3}
}

\startdata
1& {\sc kerrbb2  } & $a_* $  & $0.9985_{-0.0008}^{+0.0005} $& $ 0.9983_{-0.0037}^{+0.0005} $ & $ 0.9984_{-0.0003}^{+0.0001} $ \\
2& {\sc kerrbb2   } & $\dot{M} $  & $ 0.115  \pm 0.004 $& $ 0.115 \pm 0.007 $ & $ 0.115 \pm0.003  $\\
3& const  & -- & $1.000 \pm 0.002$& $0.999 \pm 0.002$ & $1.000 \pm 0.002$ \\
4& {\sc tbabs}  & $N_{\rm H}$ & $0.705 \pm 0.006$& $0.708 \pm 0.009$ & $0.706 \pm 0.005$ \\
5& {\sc simplr  } & $\Gamma $  & $ 2.282 \pm 0.010  $& $ 2.288 \pm 0.013 $ & $ 2.287 \pm 0.008 $\\
6& {\sc simplr  } & $f_{\rm SC} $  & $ 0.225  \pm 0.006 $& $ 0.221 \pm 0.010 $ & $ 0.221 \pm 0.005 $\\
7& {\sc kerrdisk  } & $E_{\rm L} $  & $ 6.56  \pm 0.09 $& $ 0.00  \pm 0.00  $ & $ 6.56 \pm0.05  $\\
8& {\sc kerrdisk } & $q $  & $ 2.82  \pm 0.02 $& $  2.82\pm 0.23 $ & $ 2.81 \pm 0.02 $\\
9& {\sc kerrdisk } & $N_{\rm L} $  & $ 0.015 \pm 0.001 $& $  0.000\pm 0.000 $ & $ 0.015 \pm 0.001 $\\
10& {\sc kerrdisk } & $\rm EW$ & 0.154 & -- & 0.153 \\
11& {\sc ireflect } & $\rm X_{Fe}$  & $ 5.34 \pm 0.15 $& $ 3.93 \pm 0.52 $ & $ 4.72 \pm 0.12 $\\
12& {\sc ireflect } & $ s$  & $ 0.98 \pm 0.04 $& $ 1.00 \pm 0.10 $ & $ 0.94 \pm 0.03 $\\
13& {\sc ireflect } & $\xi $  & $ 153.1 \pm 15.7 $& $ 189.6 \pm 23.7 $ & $ 153.3 \pm 11.6 $\\
\hline
14&  &$\chi^2_{\nu}$  & 1.24(561/454)&  0.82(297/361) &  ~~1.25(616/493)
\enddata

\tablenotetext{a}{The parameter set is exactly the same as for
Table~\ref{table:model_5_results}. {\bf Case 1}: Standard 0.7--45 keV
fit results for our adopted model, which are copied from column SP1 in
Table~2.  {\bf Case 2}: Fit to SPI over the range 0.7--45 keV
excluding 5.0--10.0 keV (i.e., excluding the Fe~K$\alpha$ line and Fe absorption
edge).  The value of $q$ is that returned by {\sc kerrconv}.  {\bf
Case 3}: Fit to SP1 including HEXTE data that covers the full energy
range from 0.7 to 150 keV.}
\label{table:model_5_hexte}
\end{deluxetable}

\newpage
\clearpage

\begin{appendix}

\section{A: Preliminary Nonrelativistic Analysis}

With {\sc diskbb} as the principal component, we analyzed our three
spectra in turn using three composite models of increasing
sophistication, the Models NR1--NR3 listed in Table~A1.  In addition
to {\sc diskbb}, each model includes (i) a Compton power-law
component, either {\sc compbb} \citep{nis+1986} or {\sc simplr}
\citep{ste+2009b,ste+2010b} convolved with {\sc diskbb}, and (ii) a
reflection component {\sc pexriv}~\citep{mag+1995}\footnote{As for our
adopted model, we compute only the reflected component by restricting
the fit parameter $s$ to negative values, which we here fix to -1 (see
Section~3.2).}, which models the absorption edges, plus an
Fe~K$\alpha$ emission-line feature {\sc gaussian} or {\sc
laor}~\citep{lao+1991}.  As in the case of our adopted model, also
included are the three pre-factors {\sc crabcor}, {\sc const} and {\sc
tbabs} (see Section~3.2).  In the case of Models~NR2 and NR3,
relativistic blurring effects are included using {\sc
kdblur}~\citep{lao+1991}.

{\bf Model~NR1:} \citet{cui+1998} analyzed spectrum SP1 using this
model, and we do likewise.  For the thermal component, our values of
the parameters (disk temperature $T_{\rm in}$ and inner disk radius
$R_{\rm in}$) are quite similar to those found by
\citeauthor{cui+1998}: respectively, $T_{\rm in}=(0.449\pm0.005)$ keV
(our Table~\ref{table:model_a_results}) and $T_{\rm
in}=(0.436\pm0.004)$ keV~\citep[Table 2 in][]{cui+1998}; and $R_{\rm
in}=(2.54\pm0.06)r_g$ versus $(2.68\pm0.06)r_g$ (for $D=1.86$~kpc,
$M=14.8$~\msun~and $i=27.1$~deg; $r_g\equiv GM/c^2=21.9$ km). For most
of the other fit parameters, however, the differences between our
results (Table~A2) and theirs are quite significant: For example, for
the width of the Fe~K$\alpha$ line we find $\sigma=(1.22\pm0.04)$~keV
versus $\sigma=(0.35\pm0.07)$~keV, and for the ionization parameter
$\xi\approx 0$ versus $\xi>11911$.  We attribute these differences to
our use of response files that have been significantly updated and
improved (Section~2). These new response files not only allowed us
to extend the upper energy bound to 45 keV (compared to the 30 keV
bound used by Cui et al.), but they also greatly improved the fit:
$\chi^2_{\nu} = 0.88$ versus $\chi^2_{\nu} = 1.47$. Furthermore, this
comparison is understated because Cui et al. included 1\% systematic
error in the PCU count rates whereas we used 0.5\%; for a systematic
error of 1\%, our $\chi^2_{\nu} = 0.68$.

Applying Model~NR1 to SP2 and SP3 also gives very good fits (Table A2,
columns 5 and 6).  However, comparing the results for all three
spectra one finds a wide variation in the values of the disk
temperature $T_{\rm in}$ (0.27--0.54~keV) and inner disk radius
$R_{\rm in}$ (1.95--8.30~$r_g$).  (We focus here on the parameters for
the thermal component because this component ultimately delivers our
key result, the spin of the black hole.)  We find the performance of
this model unsatisfactory for two principal reasons.  First, it
returns unlikely and near-zero values of the ionization parameter
$\xi$ (which we have here fixed to zero); this is probably because
both the reflection component and line feature are improperly modeled
(see below).  Secondly, this model implies an unrealistically strong
line feature (EW = 0.4--0.5~keV).  Therefore, we now consider an
improved model.

{\bf Model~NR2:} For the line feature, we replace the symmetric
Gaussian profile with a skewed, relativistic profile via the {\sc
laor} model, while relativistically blurring the three other additive
model components using {\sc kdblur} (Table~A1).  The {\sc laor} model
and its companion convolution model {\sc kdblur} have the serious
limitation of assuming a fixed and extreme value of the spin
parameter, $a_*=0.998$, and they therefore do not give a proper
description of black holes with moderate or low spins.  Nevertheless,
we consider these models adequate for the present purposes because
Cygnus X-1 is a rapidly spinning black hole and because the
Fe~K$\alpha$ line is only a cosmetic feature relative to the thermal
continuum (Section~4).  The merit of using {\sc laor}/{\sc kdblur}
here is that the model is simple and the values of the line strength
and ionization parameter returned by the fits are reasonable
(Table~A3).  In the end, however, as in the case of the previous
model, we find Model NR2 wanting because the parameters it returns for
the crucial thermal component ($T_{\rm in}$ and $R_{\rm in}$) differ
quite significantly for the three spectra.  We now discuss our most
physical nonrelativistic model.

{\bf Model~NR3:} In addition to our criticisms of Models NR1 and NR2,
we find these models structurally unsatisfactory because they
incorporate two different and discrepant thermal temperatures: {\sc
diskbb}'s $T_{\rm in} \sim 0.5$ keV and {\sc comptt}'s $T_{\rm 0} \sim
1.0$~keV.  In Model NR3, we solve this problem by replacing {\sc
compbb} with {\sc simplr}, which generates the Compton power-law
component via a convolution by operating on an arbitrary source of
seed photons, which in this case is the thermal component {\sc diskbb}
prior to Compton scattering.  Meanwhile, we retain the line model {\sc
laor} and the convolution model {\sc kdblur}.  Model~NR3 is specified
in Table~\ref{table:nr_models}, and our fit results are given in
Table~\ref{table:model_c_results}.  The fits are good and very
comparable to those obtained using the other two models, even though
Model~NR3 uses one less fit parameter.  We note that achieving these
fits requires iron abundances that are $\approx 3.2-5.5$ times solar.
The benefit of using this more self-consistent model is that it
harmonizes the fitted values of the parameters of the thermal
component: $T_{\rm in}=0.532,0.539~$and$~0.543$~keV and $R_{\rm
in}=2.06,2.30~$and$~2.01r_g$ for SP1, SP2 and SP3, respectively.
Meanwhile, the smallness of the inner disk radius $R_{\rm in}$ is
suggestive of the high spin revealed by our relativistic analysis (see
Section 3).

\begin{deluxetable}{ccccccccc}
\tablewidth{0pt}
\tablecaption{The three nonrelativistic models}
\tablehead{\colhead{}& \colhead{Model} 
}

\startdata
NR1 & {\sc {\small crabcor*const*tbabs(diskbb+compbb+pexriv+gaussian)}} \\
NR2 & {\sc {\small crabcor*const*tbabs(kdblur$\otimes$(diskbb+compbb+pexriv)+laor)}} \\
NR3 & ~~{\sc {\small crabcor*const*tbabs(kdblur$\otimes$(simplr$\otimes$diskbb+pexriv)+laor)}}
\enddata

\label{table:nr_models}
\end{deluxetable}

\begin{deluxetable}{cccccc}
\tablewidth{0pt}
\tablecaption{Fit results for Model NR1\tablenotemark{a}}
\tablehead{\colhead{Number}& \colhead{Model} & \colhead{Parameter}
  &\colhead{SP1}
  &\colhead{SP2} &
  \colhead{SP3}
}

\startdata
1& {\sc diskbb  } & $T_{\rm in} $  & $0.449  \pm 0.005 $& $ 0.537  \pm 0.010 $ & $0.270  \pm 0.013 $ \\ 
2& {\sc diskbb   } & $R_{\rm in} $  & $ 2.54  \pm 0.06 $& $ 1.95  \pm 0.10  $ & $ 8.30  \pm 0.98  $\\ 
3& {\sc const}  & -- & $1.012 \pm 0.002$& $1.071 \pm 0.013$ & $1.037 \pm 0.004$ \\
4& {\sc tbabs}  & $N_{\rm H}$ & $0.675 \pm 0.008$& $0.697 \pm 0.018$ & $0.862 \pm 0.022$ \\
5& {\sc gaussian  } & $E_{\rm L} $  & $ 6.40  \pm 0.14 $& $ 6.40  \pm 0.14  $ & $ 6.40 \pm0.12  $\\ 
6& {\sc gaussian } & $\sigma $  & $ 1.22  \pm 0.04 $& $  1.02\pm 0.07 $ & $ 1.02 \pm 0.04 $\\ 
7& {\sc gaussian } & $N_{\rm L} $  & $ 0.047 \pm 0.005 $& $  0.078\pm 0.011 $ & $ 0.067 \pm 0.006 $\\ 
8& {\sc gaussian } &$\rm  EW$ & 0.488 & 0.401 & 0.455 \\ 
9& {\sc compbb  } & $ T_{0} $  & $ 0.777 \pm 0.027  $& $ 0.947  \pm0.049   $ & $ 0.432 \pm 0.021 $\\ 
10& {\sc compbb  } & $ T_{\rm e} $  & $ 29.00  \pm 0.47 $& $ 21.33  \pm1.24   $ & $ 23.15 \pm 0.64 $\\ 
11& {\sc compbb  } & $\tau $  & $ 1.08 \pm 0.02  $& $ 1.24 \pm0.07  $ & $ 0.80 \pm0.05  $\\ 
12& {\sc compbb  } & $ N $  & $ 4887.9  \pm 741.6 $& $ 4074.1 \pm 978.0 $ & $ 164572.0 \pm 48271.7 $\\ 
13& {\sc pexriv } & $\rm X_{Fe} $ & $ 4.18 \pm 0.12 $& $ 2.08 \pm 0.25 $ & $ 2.29 \pm 0.08 $\\ 
14& {\sc pexriv } & $ N $  & $ 9.07 \pm 0.35 $& $ 13.17 \pm 0.84 $ & $ 14.44 \pm 0.41 $\\ 
\hline
15&  &$\chi^2_{\nu}$  & 0.88(397/453)&  1.13(326/289) &  ~~1.07(654/613)
\enddata

\tablenotetext{a}{For the model components given, the parameters from
top to bottom are: (1) inner disk temperature in keV; (2) inner disk
radius in units of $r_g\equiv GM/c^2=21.9$~km for $M=14.8M_{\odot}$;
(3) detector normalization constant relative to {\it RXTE} PCU2; (4)
hydrogen column density in units of cm$^{-2}$; (5) central line energy
in keV; (6) line width in keV; (7) line flux in units of
photons~cm$^{-2}$~s$^{-1}$; (8) equivalent width of line in keV; (9)
blackbody temperature in keV; (10) electron temperature of corona in
keV; (11) optical depth of corona; (12) normalization constant; (13)
iron abundance relative to solar; (14) normalization constant; (15)
Reduced chi-square, total chi-square and degrees of freedom,
respectively.  Details for the reflection component {\sc pexriv}:
Apart from iron, the metal abundances are solar; the photon index is
fixed at $\Gamma=2.5$, the value determined for Model~NR3; the
reflection scaling factor $s$ is fixed to -1; and the ionization
parameter is fixed, $\xi=0$ (whereas, if fitted, $\xi\approx0$).}
\tablenotetext{b}{The line central energy is pegged in the fit.}
\label{table:model_a_results}
\end{deluxetable}



\begin{deluxetable}{cccccc}
\tablewidth{0pt}
\tablecaption{Fit results for Model~NR2\tablenotemark{a}}
\tablehead{\colhead{Number}& \colhead{Model} & \colhead{Parameter}
  &\colhead{SP1}
  &\colhead{SP2} &
  \colhead{SP3}
}

\startdata
1& {\sc diskbb  } & $T_{\rm in} $  & $0.509  \pm 0.003 $& $ 0.616  \pm 0.011 $ & $0.446  \pm 0.002 $ \\ 
2& {\sc diskbb   } & $R_{\rm in} $  & $ 2.09  \pm 0.02 $& $ 1.61  \pm 0.05  $ & $ 2.40  \pm 0.01  $\\ 
3& {\sc const}  & -- & $1.014 \pm 0.002$& $1.083 \pm 0.011$ & $1.034 \pm 0.004$ \\
4& {\sc tbabs}  & $N_{\rm H}$ & $0.701 \pm 0.007$& $0.801 \pm 0.016$ & $0.717 \pm 0.006$ \\
5& {\sc laor  } & $E_{\rm L} $  & $ 6.51  \pm 0.02 $& $ 6.49  \pm 0.04  $ & $ 6.49 \pm0.02  $\\ 
6& {\sc laor } & $q $  & $ 2.37  \pm 0.07 $& $  2.54\pm 0.11 $ & $ 2.39 \pm 0.04 $\\ 
7& {\sc laor } & $N_{\rm L} $  & $ 0.012 \pm 0.001 $& $  0.035\pm 0.003 $ & $ 0.032 \pm 0.001 $\\ 
8& {\sc laor } & EW & 0.125 & 0.176 & 0.219 \\ 
9& {\sc compbb  } & $ T_{0} $  & $ 1.060 \pm 0.015  $& $ 1.279  \pm0.065   $ & $ 0.522 \pm 0.001 $\\ 
10& {\sc compbb  } & $ T_{\rm e} $  & $ 33.64  \pm 0.64 $& $ 23.23  \pm1.56   $ & $ 27.20 \pm 0.42 $\\ 
11& {\sc compbb  } & $\tau $  & $ 1.12 \pm 0.02  $& $ 1.22 \pm0.08  $ & $ 0.87 \pm0.02  $\\ 
12& {\sc compbb  } & $ N $  & $ 1404.2  \pm 83.0 $& $ 1165.7 \pm 270.6 $ & $ 59084.1 \pm 426.3 $\\ 
13& {\sc pexriv } & $\rm X_{Fe}$  & $ 2.96 \pm 0.37 $& $ 2.31 \pm 0.38 $ & $ 3.63 \pm 0.17 $\\ 
14& {\sc pexriv } & $\xi $  & $ 1168.0 \pm 115.4 $& $ 748.5 \pm 177.2 $ & $ 1647.7 \pm 115.8 $\\ 
15& {\sc pexriv } & $ N $  & $ 4.91 \pm 0.32 $& $ 14.51 \pm 0.98 $ & $ 10.33 \pm 0.35 $\\ 
\hline
16&  &$\chi^2_{\nu}$  & 1.05(474/452)&  1.16(333/288) &  ~~1.13(692/612)
\enddata

\tablenotetext{a}{Layout and parameter definitions very similar to
Table 3, with two exceptions: (1) {\sc laor} model substituted for
{\sc gaussian} and, correspondingly, the emissivity index $q$ is given
in place of the line width $\sigma$; and (2) the ionization parameter
$\xi$ is now a fit parameter rather than fixed to zero.}

\label{table:model_b_results}
\end{deluxetable}

\begin{deluxetable}{cccccc}
\tablewidth{0pt}
\tablecaption{Fit results for Model~NR3\tablenotemark{a}}
\tablehead{\colhead{Number}& \colhead{Model} & \colhead{Parameter}
  &\colhead{SP1}
  &\colhead{SP2} &
  \colhead{SP3}
}

\startdata
1& {\sc diskbb  } & $T_{\rm in} $  & $ 0.532  \pm 0.001 $& $ 0.539  \pm 0.002 $ & $ 0.543  \pm 0.001 $ \\ 
2& {\sc diskbb   } & $R_{\rm in} $  & $ 2.06  \pm 0.01 $& $ 2.30  \pm 0.01  $ & $ 2.01  \pm 0.01  $\\ 
3& {\sc const}  & -- & $1.018 \pm 0.002$& $1.078 \pm 0.010$ & $1.0295 \pm 0.0038$ \\
4& {\sc tbabs}  & $N_{\rm H}$ & $0.680 \pm 0.004$& $0.838 \pm 0.009$ & $0.699 \pm 0.004$ \\
5& {\sc simplr  } & $\Gamma $  & $ 2.206 \pm 0.011  $& $ 2.503 \pm 0.013 $ & $ 2.486 \pm 0.005 $\\ 
6& {\sc simplr  } & $f_{\rm SC} $  & $ 0.176  \pm 0.002 $& $ 0.344 \pm 0.007 $ & $ 0.317 \pm 0.003 $\\ 
7& {\sc laor  } & $E_{\rm L} $  & $ 6.56  \pm 0.02 $& $ 6.44  \pm 0.04  $ & $ 6.49 \pm0.03  $\\ 
8& {\sc laor } & $q $  & $ 2.88  \pm 0.01 $& $  2.36\pm 0.06 $ & $ 2.38 \pm 0.03 $\\ 
9& {\sc laor } & $N_{\rm L} $  & $ 0.018 \pm 0.001 $& $  0.033\pm 0.002 $ & $ 0.030 \pm 0.001 $\\ 
10& {\sc laor } & EW & 0.189 & 0.164 & 0.208 \\ 
11& {\sc pexriv } & $\rm X_{Fe}$  & $ 5.49 \pm 0.17 $& $ 3.24 \pm 0.13 $ & $ 3.22 \pm 0.14 $\\ 
12& {\sc pexriv } & $\xi $  & $ 1147.3 \pm 100.4 $& $ 906.5 \pm 81.5 $ & $ 1323.5 \pm 0.0 $\\ 
13& {\sc pexriv } & $ N $  & $ 3.78 \pm 0.29 $& $ 15.38 \pm 1.17 $ & $ 9.52 \pm 0.30 $\\ 
\hline
14&  &$\chi^2_{\nu}$  & 1.26(570/454)&  1.27(367/290) &  ~~1.12(685/614)
\enddata

\tablenotetext{a}{For definitions of most of the parameters, see
Tables~\ref{table:model_a_results} and \ref{table:model_b_results}.
The model components and parameters are the same as for
Table~\ref{table:model_b_results} with one difference: The convolution
model {\sc simplr} with two parameters has been substituted for the
additive model {\sc compbb} with four.  The two parameters of {\sc
simplr} are the photon power-law index $\Gamma$ and a normalization
constant, which is the scattering fraction $f_{\rm SC}$.}

\label{table:model_c_results}
\end{deluxetable}

\section{B: Relativistic Analysis}

{\bf Four Preliminary Models, R1--R4:} We now consider a progression
of four models that are all built around our relativistic disk model
{\sc kerrbb2} (Section~3.2).  These models progress toward our adopted
model (Section 3.2) in the sense that Model R1 is the most primitive
and our adopted model is the most physically realistic.  All four of
these models and our adopted model give very similar results for the
parameter of interest, namely $a_*$, indicating that our key result,
the extreme spin of Cygnus~X-1, is quite insensitive to the details of
the analysis.

The four preliminary models R1--R4 are defined in
Table~\ref{table:re_models}.  Every individual component in all four
models has already been described either in Section~3.2 or in 
Appendix~A.  There are only two combinations of model components that
are new and that have not been used elsewhere, namely, the reflection
components in Models R3 and R4.  Their core power-law components are
respectively {\sc compbb} and {\sc simplc}, each of which is convolved
in turn with {\sc ireflect} and {\sc kdblur}.

The results for Models~R1--R4 are presented respectively in
Tables~\ref{table:model_1_results}-\ref{table:model_4_results}.
Comparing the results for Models R1 and R2, which both employ the
widely-used reflection model {\sc pexriv}, Model~R2 is preferred
because it gives very comparable values of reduced chi-square using
two fewer fit parameters, and its reflection component {\sc simplr}
self-consistently generates the power-law with no need for the second
thermal component required by {\sc compbb}.  Furthermore, in the case
of the nonrelativistic analysis (Appendix~A), it was shown that {\sc
simplr} delivers consistent values of $R_{\rm in}$ and $T_{\rm in}$,
which {\sc compbb} fails to do.

Tables~\ref{table:model_3_results} and~\ref{table:model_4_results}
summarize our results for Models R3 and R4, which use the convolution
reflection model {\sc ireflect} in place of the additive model {\sc
pexriv}.  While both reflection models are based on the same physics,
the virtue of {\sc ireflect} is that it can take any shape of spectrum
as input, whereas {\sc pexriv} requires a power-law input spectrum.
In considering Models~R1--R4, we find Model~R4 to be superior because
it fits the data as well as the other models while using the fewest
parameters, and it is the most self-consistent (see Section~3.2). 

In a final step, as described in Section~3.2, we arrive at our adopted
model by replacing the flawed model components {\sc laor} and {\sc
kdblur} (see Appendix~A) by {\sc kerrdisk} and {\sc kerrconv}.  The
virtue of {\sc kerrdisk}/{\sc kerrconv} compared to {\sc laor}/{\sc
kdblur} is that the spin parameter is free, allowing the GR metric to
be calculated properly.  We use a single, linked value of $a_*$ in
fitting {\sc kerrdisk}/{\sc kerrconv} and {\sc kerrbb{\small 2}}.  An
inspection of Tables~B2--B5 and Table~3 shows that the key quantity,
namely the spin parameter, is precisely determined and near unity for
Models R1--R4 and our adopted model.  This indicates that the details
of how one models the power-law and reflected components has little
affect on the determination of $a_*$.

\begin{deluxetable}{ccccccccc}
\tablewidth{0pt}
\tablecaption{The four preliminary relativistic models}
\tablehead{\colhead{}& \colhead{Model} 
}

\startdata
R1 & {\sc {\small crabcor*const*tbabs(kerrbb{\small 2}+laor+kdblur$\otimes$(pexriv+compbb))}} \\
R2 & {\sc {\small crabcor*const*tbabs(simplr$\otimes$kerrbb{\small 2}+laor+kdblur$\otimes$pexriv)}} \\
R3 & {\sc {\small crabcor*const*tbabs(kerrbb{\small 2}+laor+kdblur$\otimes$(ireflect$\otimes$compbb))}} \\
R4 & ~~{\sc {\small crabcor*const*tbabs(simplr$\otimes$kerrbb{\small 2}+laor+kdblur$\otimes$(ireflect$\otimes$simplc))}}
\enddata

\label{table:re_models}
\end{deluxetable}

\begin{deluxetable}{cccccc}
\tablewidth{0pt}
\tablecaption{Fit results for Model~R1\tablenotemark{a}}
\tablehead{\colhead{Number}& \colhead{Model} & \colhead{Parameter}
  &\colhead{SP1}
  &\colhead{SP2} &
  \colhead{SP3}
}

\startdata
1& {\sc kerrbb2  } & $a_*$  & $0.9999_{-0.0001}^{+0} $& $ 0.9999_{-0.0003}^{+0} $ & $ 0.9989_{-0.0021}^{+0.0003} $ \\ 
2& {\sc kerrbb2   } & $\dot{M} $  & $ 0.102  \pm 0.007 $& $ 0.143 \pm 0.030 $ & $ 0.094 \pm0.010  $\\ 
3& const  & -- & $1.000 \pm 0.002$& $1.004 \pm 0.014$ & $1.016 \pm 0.005$ \\
4& {\sc tbabs}  & $N_{\rm H}$ & $0.760 \pm 0.008$& $0.893 \pm 0.022$ & $0.759 \pm 0.010$ \\
5& {\sc laor  } & $E_{\rm L} $  & $ 6.45  \pm 0.02 $& $ 6.42  \pm 0.04  $ & $ 6.48 \pm0.03  $\\ 
6& {\sc laor } & $q $  & $ 2.07  \pm 0.10 $& $  2.02\pm 0.20 $ & $ 2.33 \pm 0.05 $\\ 
7& {\sc laor } & $N_{\rm L} $  & $ 0.011 \pm 0.001 $& $  0.026\pm 0.003 $ & $ 0.030 \pm 0.001 $\\ 
8& {\sc laor } & EW & 0.106 & 0.131 & 0.208 \\ 
9& {\sc compbb  } & $ T_{0} $  & $ 1.023 \pm 0.005  $& $ 0.975  \pm0.028   $ & $ 0.559 \pm 0.002 $\\ 
10& {\sc compbb  } & $ T_{\rm e} $  & $ 24.23  \pm 0.47 $& $ 21.41  \pm1.06   $ & $ 24.23 \pm 0.57 $\\ 
11& {\sc compbb  } & $\tau $  & $ 1.11 \pm 0.01  $& $ 1.32 \pm0.07  $ & $ 1.05 \pm0.03  $\\ 
12& {\sc compbb  } & $ N $  & $ 1582.7  \pm 24.5 $& $ 3851.8 \pm 504.1 $ & $ 39574.0 \pm 515.0 $\\ 
13& {\sc pexriv } & $\rm X_{Fe}$  & $ 3.01 \pm 0.30 $& $ 2.87 \pm 0.38 $ & $ 3.63 \pm 0.18 $\\ 
14& {\sc pexriv } & $\xi $  & $ 751.3 \pm 104.9 $& $ 652.1 \pm 169.2 $ & $ 1226.3 \pm 0.1 $\\ 
15& {\sc pexriv } & $ N $  & $ 4.87 \pm 0.31 $& $ 12.03 \pm 1.07 $ & $ 10.87 \pm 0.40 $\\ 
\hline
16&  &$\chi^2_{\nu}$  & 1.04(470/452)&  1.26(364/288) &  ~~1.16(708/612)
\enddata

\tablenotetext{a}{For the model components given, the parameters from
top to bottom are: (1) spin parameter; (2) mass accretion rate in
units of $10^{18}$ g~s$^{-1}$; (3) detector normalization constant
relative to {\it RXTE} PCU2; (4) hydrogen column density in units of
cm$^{-2}$; (5) central line energy in keV; (6) emissivity index; (7)
line flux in units of photons~cm$^{-2}$~s$^{-1}$; (8) equivalent width
of line in keV; (9) blackbody temperature in keV; (10) electron
temperature of corona in keV; (11) optical depth of corona; (12)
normalization constant; (13) iron abundance relative to solar; (14)
ionization parameter; (15) normalization constant; (16) Reduced
chi-square, total chi-square and degrees of freedom, respectively.
Details for the reflection component {\sc pexriv}: Apart from iron,
the metal abundances are solar, and the photon index is fixed at
$\Gamma=2.5$, which is the value determined for our adopted model
(Table~\ref{table:model_5_results}).}

\label{table:model_1_results}
\end{deluxetable}

\begin{deluxetable}{cccccc}
\tablewidth{0pt}
\tablecaption{Fit results for Model~R2\tablenotemark{a}}
\tablehead{\colhead{Number}& \colhead{Model} & \colhead{Parameter}
  &\colhead{SP1}
  &\colhead{SP2} &
  \colhead{SP3}
}

\startdata
1& {\sc kerrbb2  } & $a_* $  & $0.9989_{-0.0003}^{+0.0003} $& $ 0.9998_{-0.0036}^{+0} $ & $ 0.9999_{-0.0001}^{+0} $ \\ 
2& {\sc kerrbb2   } & $\dot{M} $  & $ 0.116  \pm 0.003 $& $ 0.146 \pm 0.025 $ & $ 0.124 \pm0.010  $\\ 
3& const  & -- & $1.006 \pm 0.002$& $1.042 \pm 0.013$ & $0.972 \pm 0.004$ \\
4& {\sc tbabs}  & $N_{\rm H}$ & $0.758 \pm 0.006$& $0.841 \pm 0.018$ & $0.764 \pm 0.010$ \\
5& {\sc simplr  } & $\Gamma $  & $ 2.295 \pm 0.011  $& $ 2.507 \pm 0.012 $ & $ 2.525 \pm 0.008 $\\ 
6& {\sc simplr  } & $f_{\rm SC} $  & $ 0.179  \pm 0.002 $& $ 0.328 \pm 0.009 $ & $ 0.299 \pm 0.003 $\\ 
7& {\sc laor  } & $E_{\rm L} $  & $ 6.54  \pm 0.03 $& $ 6.41  \pm 0.04  $ & $ 6.51 \pm0.03  $\\ 
8& {\sc laor } & $q $  & $ 2.87  \pm 0.02 $& $  2.27\pm 0.07 $ & $ 2.41 \pm 0.05 $\\ 
9& {\sc laor } & $N_{\rm L} $  & $ 0.014 \pm 0.001 $& $  0.032\pm 0.002 $ & $ 0.028 \pm 0.001 $\\ 
10& {\sc laor } & $\rm EW$ & 0.148 & 0.154 & 0.187 \\ 
11& {\sc pexriv } &$\rm X_{Fe}$  & $ 6.23 \pm 0.17 $& $ 3.06 \pm 0.12 $ & $ 3.60 \pm 0.15 $\\ 
12& {\sc pexriv } & $\xi $  & $ 726.4 \pm 39.5 $& $ 981.7 \pm 103.4 $ & $ 1163.5 \pm 73.4 $\\ 
13& {\sc pexriv } & $ N $  & $ 6.91 \pm 0.44 $& $ 15.23 \pm 1.01 $ & $ 12.28 \pm 0.53 $\\ 
\hline
14&  &$\chi^2_{\nu}$  & 1.24(561/454)&  1.28(371/290) &  ~~1.18(725/614)
\enddata

\tablenotetext{a}{For definitions of most of the parameters, see
Table~\ref{table:model_1_results}.  Two distinctions between this
table and Table~\ref{table:model_1_results}: (1) The convolution model
{\sc simplr} with two parameters has been substituted for the additive
model {\sc compbb} with four.  The two parameters of {\sc simplr} are
the photon power-law index $\Gamma$ and a normalization constant,
which is the scattering fraction $f_{\rm SC}$.  (2) The photon index
in {\sc pexriv} is not fixed, rather it is linked to the photon index
in {\sc simplr}.}

\label{table:model_2_results}
\end{deluxetable}

\begin{deluxetable}{cccccc}
\tablewidth{0pt}
\tablecaption{Fit results for Model~R3\tablenotemark{a}}
\tablehead{\colhead{Number}& \colhead{Model} & \colhead{Parameter}
  &\colhead{SP1}
  &\colhead{SP2} &
  \colhead{SP3}
}

\startdata
1& {\sc kerrbb2  } & $a_* $  & $0.9989_{-0.0007}^{+0.0003} $& $ 0.9999_{-0.0003}^{+0} $ & $ 0.9888_{-0.0066}^{+0.0037} $ \\ 
2& {\sc kerrbb2   } & $\dot{M} $  & $ 0.110  \pm 0.004 $& $ 0.145 \pm 0.031 $ & $ 0.098 \pm0.008  $\\ 
3& const  & -- & $1.001 \pm 0.002$& $0.984 \pm 0.015$ & $1.026 \pm 0.004$ \\
4& {\sc tbabs}  & $N_{\rm H}$ & $0.733 \pm 0.006$& $0.820 \pm 0.025$ & $0.670 \pm 0.011$ \\
5& {\sc laor  } & $E_{\rm L} $  & $ 6.43  \pm 0.02 $& $ 6.42  \pm 0.04  $ & $ 6.45 \pm0.03  $\\ 
6& {\sc laor } & $q $  & $ 1.78  \pm 0.14 $& $  1.83\pm 0.28 $ & $ 2.26 \pm 0.06 $\\ 
7& {\sc laor } & $N_{\rm L} $  & $ 0.011 \pm 0.001 $& $  0.024\pm 0.003 $ & $ 0.030 \pm 0.001 $\\ 
8& {\sc laor } & $\rm EW$ & 0.109 & 0.117 & 0.204 \\ 
9& {\sc compbb  } & $ T_{0} $  & $ 0.942 \pm 0.010  $& $ 1.004  \pm0.017   $ & $ 0.518 \pm 0.001 $\\ 
10& {\sc compbb  } & $ T_{\rm e} $  & $ 26.25  \pm 0.43 $& $ 22.85  \pm0.61   $ & $ 26.25 \pm 0.48 $\\ 
11& {\sc compbb  } & $\tau $  & $ 1.19 \pm 0.02  $& $ 1.24 \pm0.04  $ & $ 0.90 \pm0.03  $\\ 
12& {\sc compbb  } & $ N $  & $ 2212.8  \pm 99.1 $& $ 3301.5 \pm 234.5 $ & $ 59300.0 \pm 532.3 $\\ 
13& {\sc ireflect } & $\rm X_{Fe}$  & $ 2.20 \pm 0.19 $& $ 2.02 \pm 0.26 $ & $ 3.51 \pm 0.18 $\\ 
14& {\sc ireflect } & $ s$  & $ 0.36 \pm 0.02 $& $ 0.72 \pm 0.03 $ & $ 0.72 \pm 0.03 $\\ 
15& {\sc ireflect } & $\xi $  & $ 84.2 \pm 9.0 $& $ 59.2 \pm 12.5 $ & $ 65.5 \pm 6.8 $\\ 
\hline
16&  &$\chi^2_{\nu}$  & 1.07(483/452)&  1.25(358/288) &  ~~1.14(695/612)
\enddata

\tablenotetext{a}{Here, we have substituted the convolution model {\sc
ireflect} for the additive reflection model {\sc pexriv}.  These
models have two parameters in common, the iron abundance and
ionization parameter, and one that differs, namely, the normalization
of {\sc pexriv} is replaced by the reflection scaling factor $s$.  For
the definitions of all other parameters, see
Tables~\ref{table:model_1_results} and \ref{table:model_2_results}.}

\tablenotetext{b}{In applying {\sc ireflect}, the reflection scaling
factor $s$ is negative; here we give absolute values of $s$.}

\label{table:model_3_results}
\end{deluxetable}

\begin{deluxetable}{cccccc}
\tablewidth{0pt}
\tablecaption{Fit results for Model~R4\tablenotemark{a}}
\tablehead{\colhead{Number}& \colhead{Model} & \colhead{Parameter}
  &\colhead{SP1}
  &\colhead{SP2} &
  \colhead{SP3}
}

\startdata
1& {\sc kerrbb2  } & $a_* $  & $0.9987_{-0.0005}^{+0.0004} $& $ 0.9997_{-0.0026}^{+0.0001} $ & $ 0.9999_{-0.0001}^{+0} $ \\
2& {\sc kerrbb2   } & $\dot{M} $  & $ 0.115  \pm 0.004 $& $ 0.143 \pm 0.026 $ & $ 0.122 \pm0.011  $\\
3& const  & -- & $1.000 \pm 0.002$& $1.037 \pm 0.013$ & $0.726 \pm 0.003$ \\
4& {\sc tbabs}  & $N_{\rm H}$ & $0.704 \pm 0.006$& $0.761 \pm 0.022$ & $0.687 \pm 0.010$ \\
5& {\sc simplr  } & $\Gamma $  & $ 2.284 \pm 0.010  $& $ 2.525 \pm 0.012 $ & $ 2.551 \pm 0.011 $\\
6& {\sc simplr  } & $f_{\rm SC} $  & $ 0.227  \pm 0.006 $& $ 0.329 \pm 0.013 $ & $ 0.308 \pm 0.006 $\\
7& {\sc laor  } & $E_{\rm L} $  & $ 6.54  \pm 0.03 $& $ 6.42  \pm 0.04  $ & $ 6.48 \pm0.04  $\\
8& {\sc laor } & $q $  & $ 2.85  \pm 0.02 $& $  2.26\pm 0.07 $ & $ 2.33 \pm 0.07 $\\
9& {\sc laor } & $N_{\rm L} $  & $ 0.016 \pm 0.001 $& $  0.030\pm 0.002 $ & $ 0.025 \pm 0.002 $\\
10& {\sc laor } & EW & 0.161 & 0.150 & 0.165 \\
11& {\sc ireflect } & $\rm  X_{Fe}$  & $ 5.33 \pm 0.15 $& $ 3.16 \pm 0.12 $ & $ 3.63 \pm 0.14 $\\
12& {\sc ireflect } & $ s$  & $ 1.00 \pm 0.04 $& $ 0.89 \pm 0.03 $ & $ 0.95 \pm 0.03 $\\
13& {\sc ireflect } & $\xi $  & $ 148.0 \pm 14.7 $& $ 53.6 \pm 7.9 $ & $ 60.8 \pm 8.2 $\\
\hline
14&  &$\chi^2_{\nu}$  & 1.20(545/454)&  1.28(370/290) &  ~~1.17(720/614)
\enddata

\tablenotetext{a}{Same model components and parameters as in
Table~\ref{table:model_3_results}, except that we replace the
power-law model {\sc compbb} with the convolution model {\sc simplr},
thereby reducing the four parameters of the former model component to
two, namely the photon power-law index $\Gamma$ and the normalization
constant $f_{\rm SC}$.}

\label{table:model_4_results}
\end{deluxetable}

\end{appendix}

\end{document}